\documentclass[11pt]{article}
\usepackage{jheppub}
\usepackage{axodraw2,pix}
\usepackage{epsfig}
\usepackage{amsfonts}
\usepackage{amsmath}
\usepackage{amssymb}
\usepackage{bbm,bm}
\usepackage{slashed}

\usepackage{tikz}
\newdimen\nodeDist
\allowdisplaybreaks[1]

\renewcommand{\vec}[1]{{\bf #1}}
\newcommand{\rmii}[1]{{\mbox{\tiny\rm{#1}}}}
\newcommand{\gammaE}{{\gamma_\rmii{E}}}
\newcommand{\bmu}{\bar\mu}
\newcommand{\rmO}{{\mathcal{O}}}

\newcommand{\mD}{m_\rmii{D}}

\newcommand{\gE}{g_\rmii{3d}}

\newcommand{\lambdaE}{\lambda_\rmii{E}}
\newcommand{\Tc}{T_{\rm c}}

\newcommand{\xperp}{x_{\perp}}

\newcommand{\Tint}[1]{{\hbox{$\sum$}\!\!\!\!\!\!\!\int\,}_{\!\!\!\!\raise-0.9ex\hbox{$\scriptstyle{#1}$}}}
\newcommand{\Tinti}[1]{{{\Sigma}\!\!\!\!\raise0.3ex\hbox{$\int$}_\rmii{${#1}$}}}
\newcommand{\Tintip}[1]{{{\Sigma'}\!\!\!\!\!\raise0.3ex\hbox{$\int$}_\rmii{${#1}$}}}

\def\OO{\mathcal{O}}

\def\bop{{\boldsymbol p}}

\def\bk{{\boldsymbol k}}
\def\bq{{\boldsymbol q}}

\def\alphas{\alpha_{\mathrm{s}}}

\def\gsqa{g_{\rmii{3d}}^2a}

\def\gsql{g_{\rmii{3d}}^2 L}

\def\gsqlmin{g_{\rmii{3d}}^2 L_\rmii{min}}
\def\gsq{g_{\rmii{3d}}^2}
\def\gfour{g_{\rmii{3d}}^4}
\def\gsix{g_{\rmii{3d}}^6}

\def\mDsq{\mD^2}

\def\mIsq{m_{\infty}^2}
\def\mIsqq{m_{\infty,{\rm q}}^2}
\def\mIsqg{m_{\infty,{\rm g}}^2}

\def\Tr{\mathrm{Tr}}

\def\EE{\langle EE \rangle}
\def\BB{\langle BB \rangle}
\def\EB{i \langle EB \rangle}
\def\xc{x_\mathrm{cont}}
\def\yc{y_\mathrm{cont}}

\def\Zg{Z_{\rm g}}
\def\Zgtd{\Zg^{\rmii{3d}}}
\def\Zf{Z_{\rm f}}
\def\ZP{Z_\rmii{P}}
\def\ZEE{Z_\rmii{E,E}}
\def\ZEB{Z_\rmii{E,B}}
\def\ZBE{Z_\rmii{B,E}}
\def\ZBB{Z_\rmii{B,B}}

\def\CA{C_\rmii{A}}
\def\CF{C_\rmii{F}}
\def\Nc{N_\mathrm{c}}
\def\Nf{N_\rmi{f}}
\def\Tf{T_\rmii{F}}
\def\dA{d_\rmii{A}}
\def\dR{d_\rmii{R}}
\def\e{\mathrm{e}}
\def\ge{\gamma_\mathrm{E}}

\def\TopoWR(#1,#2){\;\pic{
  #1(0,15)(30,15)
  #2(15,15)(15,0,180)%
  \GCirc(0,15){2}{0.75}
  \GCirc(30,15){2}{0.75}
 }}
\def\ToptWBB(#1,#2,#3){\;\pic{
  #1(0,15)(30,15)
  #2(15,15)(15,0,180)%
  #3(15,15)(7.5,0,180)%
  \GCirc(0,15){2}{0.75}
  \GCirc(30,15){2}{0.75}
  \Vertex(7.5,15){1}
  \Vertex(22.5,15){1}
 }}
\def\ToptWSBB(#1,#2,#3,#4,#5){\;\pic{
  #1(0,15)(30,15)
  #2(15,15)(15,0,70)
  #3(15,15)(15,110,180)
  \GCirc(15,15 70 sin 15 mul add){5}{0.75}
  #4(15,15 70 sin 15 mul add)(5,0,180)
  #5(15,15 70 sin 15 mul add)(5,180,360)%
  \GCirc(0,15){2}{0.75}
  \GCirc(30,15){2}{0.75}
  }}
\def\ToptWSBSB(#1,#2,#3){\;\pic{
  #1(0,15)(30,15)
  #2(6,15)(6,0,180)%
  #3(24,15)(6,0,180)%
  \GCirc(0,15){2}{0.75}
  \GCirc(30,15){2}{0.75}
  \Vertex(12,15){1}
  \Vertex(18,15){1}
  }}
\def\ToptWSBBS(#1,#2,#3){\;\pic{
  #1(0,15)(30,15)
  #2(10,15)(10,0,180)%
  #3(20,15)(10,180,360)%
  \GCirc(0,15){2}{0.75}
  \GCirc(30,15){2}{0.75}
  \Vertex(10,15){1}
  \Vertex(20,15){1}
  }}
\def\ToptWS(#1,#2,#3){\;\pic{
  #1(30,15)(0,15)
  #2(15,15)(15,0,180)
  #3(15,15)(15,180,360)%
  \GCirc(0,15){2}{0.75}
  \GCirc(30,15){2}{0.75}
  }}
\def\ToptWM(#1,#2,#3,#4){\;\pic{
  #1(0,15)(30,15)
  #2(15,15)(15,0,90)
  #3(15,15)(15,90,180)
  #4(15,30)(15,15)%
  \GCirc(0,15){2}{0.75}
  \GCirc(30,15){2}{0.75}
  \Vertex(15,15){1}
  }}
\def\ToptWSal(#1,#2,#3,#4){\;\pic{
  #1(0,15)(30,15)
  #2(15,15)(15,0,90)
  #3(15,15)(15,90,180)
  #4(0,30)(15,270,360)
  \GCirc(0,15){2}{0.75}
  \GCirc(30,15){2}{0.75}
  }}
\def\ToptWal(#1,#2,#3){\;\pic{
  #1(0,15)(30,15)
  #2(15,15)(15,0,180)
  #3(7.5,15)(7.5,0,180)%
  \GCirc(0,15){2}{0.75}
  \GCirc(30,15){2}{0.75}
  \Vertex(15,15){1}
  }}
\def\ToptWar(#1,#2,#3){\;\pic{
  #1(0,15)(30,15)
  #2(15,15)(15,0,180)
  #3(22.5,15)(7.5,0,180)%
  \GCirc(0,15){2}{0.75}
  \GCirc(30,15){2}{0.75}
  \Vertex(15,15){1}
  }}
\def\ToptWTT(#1,#2,#3){\;\pic{
  #1(0,15)(30,15)
  #2(0,22.5)(7.5,-90,270)%
  #3(30,22.5)(7.5,-90,270)%
  \GCirc(0,15){2}{0.75}
  \GCirc(30,15){2}{0.75}
 }}
\def\ToptWTR(#1,#2,#3){\;\pic{
  #1(0,15)(30,15)
  #2(0,22.5)(7.5,-90,270)%
  #3(22.5,15)(7.5,0,180)%
  \GCirc(0,15){2}{0.75}
  \GCirc(30,15){2}{0.75}
  \Vertex(15,15){1}
 }}

\def\scfc{0.7}  
\def\phgt{21}   
\def\pwcb{31.5} 
\def\pwcc{42} 
\newcommand{\PIC}[4]{\;\parbox[c]{#2 pt}{\begin{picture}(#2,#3)(0,0)
\SetWidth{1.0}\SetScale{#4} #1 \end{picture}}\;}
\renewcommand{\picb}[1]{\PIC{#1}{\pwcb}{\phgt}{\scfc}}
\renewcommand{\picc}[1]{\PIC{#1}{\pwcc}{\phgt}{\scfc}}

\title{The force-force-correlator in hot QCD perturbatively and from the lattice}
\author[a]{Jacopo Ghiglieri,}
\author[b]{Guy D.\ Moore,}
\author[c]{Philipp Schicho,}
\author[c,d]{Niels Schlusser}

\affiliation[a]{
SUBATECH, Universit\'e de Nantes, IMT Atlantique, IN2P3/CNRS,\\
4 rue Alfred Kastler,
La Chantrerie BP 20722, 44307 Nantes, France}
\affiliation[b]{
Institut f\"ur Kernphysik, Technische Universit\"at Darmstadt,\\
Schlossgartenstrasse 2,
D-64289 Darmstadt, Germany}
\affiliation[c]{
Department of Physics and Helsinki Institute of Physics,\\
P.O.~Box 64, FI-00014 University of Helsinki, Finland}
\affiliation[d]{
Biozentrum, University of Basel,\\
4056 Basel, Switzerland}
\emailAdd{jacopo.ghiglieri@subatech.in2p3.fr}
\emailAdd{guy.moore@physik.tu-darmstadt.de}
\emailAdd{schicho@itp.uni-frankfurt.de}
\emailAdd{niels.schlusser@unibas.ch}

\abstract{
High-energy particles traversing a medium experience modified dispersion.
In the Quark-Gluon Plasma, such dispersion
affects jet propagation and transport properties and
should be determined better.
Above $\sim\!2\Tc$ we expect
strongly coupled infrared behavior and
perturbative ultraviolet behavior, allowing a perturbative matching to
an effective theory called EQCD, which can be studied non-perturbatively.
We study the relevant non-local operator in EQCD at next-to-leading order which 
allows for a complete EQCD-to-lattice match and
prepares the groundwork for a matching between EQCD and full QCD. 
Our results in EQCD show remarkable agreement between perturbation 
theory and the lattice in the expected regime.
}


\preprint{HIP-2021-43/TH}

\begin{document}
\maketitle

%
\section{Introduction}
\label{sec:intro}
The Quark-Gluon Plasma (QGP), an exotic state of strongly
interacting matter, is currently investigated in heavy-ion collision 
experiments. Its experimental characterization and the accompanying theoretical 
activity proceed along the axes of bulk properties and hard probes. The former
is the study of the behavior of the many lower-energy produced particles,
which are understood to arise from the hadronization of a hydrodynamically-evolving, near-equilibrium
medium, while the latter concentrates on the few particles which are very energetic or weakly coupled to the medium.

Jets, a key hard probe, are important experimental sources of evidence about the nature of the strong nuclear interaction under extreme conditions --
see \cite{Connors:2017ptx,Cunqueiro:2021wls} for recent reviews.
Jets are generated by colored particles at high energies,
a regime where the theory of the strong nuclear force, Quantum Chromodynamics (QCD), is supposed to be weakly coupled and accessible to perturbative methods.
Here we focus on jets created by light constituents such as
light quarks ($u,d,s$) or gluons.
It was found by
Klimov~\cite{Klimov:1981ka,Klimov:1982bv} and independently by
Weldon~\cite{Weldon:1982aq,Weldon:1982bn}
that -- though being massless -- high-energy particles with momentum $\bk$ follow the dispersion relation of massive particles when traversing the QGP
\begin{equation}
\label{massive_dispersion}
  \omega_\bk^2 = \bk^2 + m^2_\bk
\;.
\end{equation}
By scattering with the medium they acquire an effective mass, which,
at very large momentum, is called the {\em asymptotic mass}. 
To one-loop order these masses are composed of
the gauge condensate $\Zg$ and
the fermion condensate $\Zf$:
\begin{equation}
\label{eff_mass}
\mIsqq = g^2\CF^{ }\left( \Zg^{ } + \Zf^{ } \right)
\,, \qquad
\mIsqg = g^2\CA^{ }\Zg^{ } + 2g^2 \Tf^{ }\Nf^{ }\Zf^{ }
\,,
\end{equation}
where
$\mIsqq$ applies for quarks and
$\mIsqg$ applies for gluons.
Here
$\CF=(\Nc^2-1)/(2\Nc)$ is the quadratic Casimir of the quark representation,
$\CA=\Nc$ is the adjoint Casimir,
$\Nf$ is the number of light (Dirac) quark species, and
$\Tf=1/2$ is the Dynkin index for the quarks.
The condensates $\Zg$ and $\Zf$
are non-local and
have a gauge-invariant definition%
\footnote{
Eq.~\eqref{eff_mass} can be seen as arising from
integrating out the energy scale of the jet $E\gg T$ and
truncating at first order in $T/E$, as well as determining
the matching coefficients at first order in $g$.
If the scale of the hard parton is $E\gtrsim T$ rather than $E\gg T$,
higher orders in the $T/E$ expansion become relevant.
These additional operators could spoil
the factorization into fermionic and bosonic condensates as of~\eqref{eff_mass}.
We will return to this issue
when discussing the contribution of the scale $T$ at higher orders in an upcoming paper.
}
in terms of correlators~\cite{Braaten:1991gm,CaronHuot:2008uw}
\begin{align}
\label{mom_ferm_cond}
  \Zf &\equiv \frac{1}{2\dR} \Big\langle
    \overline{\psi} \frac{v_\mu\gamma^\mu}{v\cdot D} \psi
  \Big\rangle
  \;,\\
\label{mom_gauge_cond}
  \Zg &\equiv - \frac{1}{\dA} \Big\langle
    v_{\alpha}^{ } F^{\alpha\mu}
    \frac{1}{(v \cdot D)^2}
    v_{\nu}^{ } F^\nu_{\;\; \mu}
  \Big\rangle
  \;,
\end{align}
where $v^{\mu}=(1,\vec{v})$ is the light-like four-velocity of the hard particle,
$d_{\rmii{R,A}}$ are the dimensions
of the fermion and adjoint representations respectively,
and the expectation value $\langle\dots\rangle$ denotes a thermal expectation value.
Our conventions are that the metric is the mostly-plus one,
the covariant derivative is
$D_\mu = \partial_\mu - i g A_\mu$ (with the gauge coupling $g$) and
the field strength tensor is
$F^{\mu\nu} = \frac{i}{g} \left[ D^\mu , D^\nu \right]$.
Since we are interested in QCD, henceforth we will specialize to the gauge group SU(3),
with
$\CF=4/3$,
$\CA=3$,
$d_{\rmii{R}}=3$,
$d_{\rmii{A}}=8$, and
$\Tf=1/2$ where not indicated differently.

Eq.~\eqref{mom_gauge_cond} can be rewritten in coordinate space, where the inverse powers of derivatives describe an integral over separations, with a Wilson line to reflect that they are covariant derivatives.  That is,
$\Zg$ is an integral over
the lightlike separation
$x^+\equiv(x^0+{\bf v}\cdot {\bf x})/2$
of a correlator of two covariant Lorentz-force insertions
\begin{equation}
\label{int_gauge_corr}
  \Zg = - \frac{1}{\dA} \int_0^\infty \! {\rm d} x^+ x^+
  \Bigl\langle
    v_\mu^{ } F^{\mu \nu}_a (x^+)\,U^{ab}_\rmii{A}(x^+;0)\,
    v_\rho^{ } F^\rho_{b \, \nu}(0)
  \Bigr\rangle
  \;,
\end{equation}
where $U_\rmii{A}(x^+;0)$ is an
adjoint, lightlike Wilson line.%
\footnote{
  The Wilson line structure of~\eqref{int_gauge_corr} is a somewhat delicate issue that requires further clarification.
  Group-theoretically, an adjoint line is equivalent
  to a pair of fundamental Wilson lines with the same endpoints, but they may differ in terms of operator ordering or endpoints, as the fundamental lines may stretch back to $x^+=-\infty$.
  If we treat $v = 1 + \epsilon$ then all operators along the Wilson line commute, and there
  is no difference between the treatments.  But if collinear physics becomes important at
  some higher order in the 4-dimensional treatment, then there will be a discontinuity
  between
  a $v=1+\epsilon$ and
  a $v=1-\epsilon$ treatment and the distinction
  between Wilson line types may become important.
  This cannot occur in eq.~\eqref{Zg_EQCD}, the EQCD version of eq.~\eqref{int_gauge_corr} which we study in this paper, because
  of the time-independence of EQCD. We will come back to this 
  issue when addressing the $T$-scale contribution at $\OO(g^2)$ in
  an upcoming paper.
}

Despite the jet momentum being much larger than
the temperature of the medium that it traverses,
the interaction between jet and medium still receives contributions from
the infrared (IR) regime, i.e.\ from
energy-momentum regions of $\mathcal{O}(gT)$ or smaller.
These contributions can, for many quantities such as
the interaction rate, be dominant. Indeed, 
the emergence of these IR scales causes 
the perturbative series to be an expansion in $g$ rather than in $\alphas{}$; furthermore,  at an operator-dependent order in $g$, the perturbative expansion breaks
down, due to the presence of the so-called \emph{magnetic} or non-perturbative
scale $g^2T$ \cite{Linde:1980ts}. 
In our case, the IR contributions affect the two operators at different orders.
The fermionic condensate $\Zf$ is well under perturbative control, with no IR 
contribution appearing as an $\OO{(g)}$ correction to
the leading order (LO) result~\cite{CaronHuot:2008uw}
\begin{equation}
  \Zf^{ }
    = \Zf^\rmii{LO} + \delta\Zf^{ }
    = \frac{T^{2}}{12} + 0
    + \rmO(g^2)
  \,.
\end{equation}
Only at higher orders it receives contributions from the IR regime.
However, the gauge condensate $\Zg$ receives IR contributions 
already at $\OO(g)$: the next-to-leading order (NLO) result reads~\cite{CaronHuot:2008uw}
\begin{equation}
\label{eq:NLO_Zg}
  \Zg^{ }
  = \Zg^\rmii{LO} + \delta\Zg^{ }
    = \frac{T^{2}}{6} - \frac{T \mD}{2\pi}
    + \rmO(g^2)
  \,,
\end{equation}
where at leading order the Debye screening mass
$\mD^{2}$ is given by~\cite{Weldon:1982aq}
\begin{equation}
  \mDsq =
    \frac{g^2 T^2}{3} \left( \CA + \Tf \Nf \right)
    + \OO(g^4)
  \;.
\end{equation}
Formally $\mDsq$ is a parameter of an infrared
effective description, EQCD, which we will introduce momentarily.
At leading order it equals twice the gluonic asymptotic mass,
$\mDsq = 2 \mIsqg$.
For the values of $T$ and therefore $g$ which are relevant
in any conceivable heavy-ion experiment,
the Debye mass is large enough for the two known terms in \eqref{eq:NLO_Zg} to be comparable.
Therefore, the 
IR contribution spoils
the convergence of the perturbative expansion.

In a remarkable set of papers~\cite{CaronHuot:2008ni,CaronHuot:2008uw}
it was however shown that the leading
soft interactions of a very hard, hence light-like, parton with a soft thermal
bath can be greatly simplified, both for
the interaction rate~\cite{CaronHuot:2008ni} and for
$\Zg$ \cite{CaronHuot:2008uw}.
In more detail, they can be isolated and treated in a 
dimensionally-reduced 
Effective Field Theory of thermal QCD, called {\em Electrostatic QCD} (EQCD)~\cite{Braaten:1994na,Braaten:1995cm,Braaten:1995jr,Kajantie:1995dw,Kajantie:1997tt}.
Its effective parameters encode the ultraviolet (UV)
behavior of thermal QCD as a function
of temperature $T$ and number of massless quark flavors $\Nf$.
This is achieved by matching the Green's functions of both theories in the IR.
To investigate a QCD operator using this effective field theory, we must determine the EQCD counterpart of that operator; for us, this means determining the EQCD equivalent of the correlator of eq.~\eqref{int_gauge_corr}.

Our long-term goal, started in \cite{Moore:2020wvy}, is to determine $\Zg$
by a matching between full QCD and EQCD and
by computing the EQCD contribution
non-perturbatively on the lattice.
This paper will take two steps
in this direction, leaving one step for future work.
First, we will
update the lattice part of the EQCD equivalent to eq.~\eqref{int_gauge_corr}. 
Second, we will
calculate the EQCD correlator to next-to-leading order in EQCD perturbation theory.
This improves the fitting necessary in
the lattice determination, and it also sets the groundwork
for the matching to full QCD, by
gaining a better analytical understanding of the (unphysical) UV
divergences in EQCD, and how these divergences are appropriately dealt with.
We argue on dimensional grounds that NLO is the last order at which potentially
UV-divergent terms from EQCD in~\eqref{int_gauge_corr} can arise.
When supplied with a full four-dimensional perturbative calculation at the same order, 
the matching of our lattice EQCD results back to full QCD would be complete, and
entirely IR-resummed values for $\mIsq$ can be provided.
The four-dimensional calculation is spared for a separate publication, however.
Moreover, one can finally only interpret
the UV divergences in EQCD in the context of the full four-dimensional framework.
Therefore, a fully consistent subtraction scheme also has to be postponed to another publication.

This paper is organized as follows:
Section~\ref{sec:power:counting} clarifies the power counting scheme(s) used
throughout this present publication.
Section~\ref{sec:transp_from_DR} details the transition from full QCD to EQCD.
We present the analytical calculation of the force-force correlator in EQCD at NLO
in sec.~\ref{sec:matching} and 
update the existing lattice calculation
in sec.~\ref{sec:latt}.
Section~\ref{sec:discussion}
discusses our results with an outlook to future interesting research. 
Details of our perturbative calculation are found in appendix~\ref{sec:NLO:details}.

%
\section{Power-counting scheme}
\label{sec:power:counting}
To shed light on the at first sight opaque power counting scheme of our result, we find it instructive to parametrize our result before actually diving into the computation.
Let us start with a power-counting analysis
of $\Zg$. Its leading order form is 
\begin{equation}
\label{zglo}
  \Zg^{\rmii{LO}} = 2\int\frac{{\rm d}^3p}{(2\pi)^3}\frac{n_{\rmii{B}}(p)}{p}
  \;,
\end{equation}
where $n_\rmii{B}$ is the Bose--Einstein distribution.
This form shows how
the LO term in eq.~\eqref{eq:NLO_Zg} is obtained: 
it is the leading contribution from the momentum region $p\sim T$. 
The first IR contribution arises from $p\sim gT$, which
is {\em Bose-enhanced}, i.e.\
$n_{\rmii{B}}(p)\approx T/p\sim 1/g$.
This, together with ${\rm d}^3p/p\sim g^2T^2$, suggests
an $\OO(g)$ contribution from this scale, corresponding to the 
contribution to $\Zg$ from the leading order in EQCD.
Going further to the IR, we can expect the first contribution from
the non-perturbative scale $g^2T$ at $\OO(g^2)$.
This is precisely the order reached by our NLO-in-EQCD perturbative
calculation, as well as that of the second-order contribution from the scale $p\sim T$.
Hence, a strict perturbative expansion of $\Zg$ in the QCD coupling $g$ would yield 
\begin{align}
  &
  \hphantom{\bigg[\quad}
  \mathrm{scale\;}T
  &&
  \mathrm{scale\;}gT
  &&
  \mathrm{scale\;}g^2T
  &
  \nn
  \Zg =
    &\bigg[\quad
      \frac{T^2}{6}
    - \frac{T\mu_\rmi{h}^{ }}{\pi^2}
    &&
    &&
    &\bigg]
  \label{loline}\\
  + &\bigg[\quad
    &&
    - \frac{T\mD}{2\pi}
    + \frac{T\mu_\rmi{h}^{ }}{\pi^2}
    &&
  &\bigg]
  \label{nloline}\\
  + &\bigg[\quad
      c^{\ln}_{\rmi{hard}} \ln\frac{T}{\mu_\rmi{h}}
    + c_{\rmii{$T$}}^{ }
    &&
    + c^{\ln}_{\rmi{hard}} \ln\frac{\mu_\rmi{h}}{\mD} 
    + c^{\ln}_{\rmi{soft}} \ln\frac{\mD}{\mu_\rmi{s}}
    + c_{\rmii{$gT$}}^{ }
    &&
    + c^{\ln}_{\rmi{soft}} \ln\frac{\mu_\rmi{s}}{g^2T}
    + c_{\rmii{$gT^2$}}^{ }
  &\bigg]
  \label{nnloline}\\[2mm]
  +&\,\OO(g^3)\;,&
  &&
\end{align} 
where the three lines correspond to
$\OO(g^{0})$ -- $\OO(g^{2})$, respectively, whereas
columns correspond to the originating scale.
At $\mathcal{O}(g^2)$ all scales contribute, and the contribution of each of them is 
\emph{scheme-dependent}: we indicate this through the 
intermediate regulators
$T\gg \mu_\rmi{h}\gg gT$ and
$gT\gg \mu_\rmi{s}\gg g^2T$.%
\footnote{\label{foot_lo_div}
  The zeroth- and first-order terms in $g$ are also in principle regulator-dependent.
  This corresponds to the linear UV divergence of the LO-in-EQCD result, which
  is in general dealt with through a simple subtraction scheme, so that
  a cutoff does not appear explicitly.
}
These should not be taken as
indicating the choice of a cutoff scheme, but rather as generic placeholders
for an arbitrary choice of scheme to separate the contribution of the scales.
This scheme dependence also affects the coefficients
$c_{\rmii{$T$}}$,
$c_{\rmii{$gT$}}$ and
$c_{\rmii{$g^2T$}}$,
with the latter
being non-perturbative.
Through our calculation we will be able to determine
$c^{\ln}_{\rmi{hard}}$ and
$c^{\ln}_{\rmi{soft}}$ from the logarithmic
divergences of the NLO-in-EQCD calculation.\footnote{%
The form of eq.~\eqref{nnloline} is to be taken as a sketch,
derived from dimensional considerations only,
which suggest the presence of a logarithmic sensitivity
between the scales. Our explicit evaluation finds
a single- and
double-logarithmic sensitivity between the $g^2T$ and
$gT$ scales.}
That is not, however,
our main aim, which is instead to incorporate the all-order EQCD contribution
to $\Zg$ through lattice EQCD.

Finally, we remark that, due to the super-renormalizable nature of EQCD,
the EQCD contributions of $\mathcal{O}(g^3)$ and higher will not
present further UV divergences.
In the language of this section, we can expect the UV behavior
at $\mathcal{O}(g^n)$ to be $\mu_h^{2-n}$, so that, after the
leading linear and
subleading log divergences we can only find negative power laws in the UV cutoff.

%
\section{Light-cone observables and dimensionally-reduced theories}
\label{sec:transp_from_DR}

As we noted, it was found by Caron-Huot that the infrared,
non-perturbative part of the jet-medium interaction can be isolated in
the framework of electrostatic QCD (EQCD)~\cite{CaronHuot:2008ni}.
Since the gluonic Matsubara zero-mode contributes a factor of $1/g$ for each closed thermal loop,
the perturbative power counting scheme of a $g^2$-suppression per loop is spoiled.
In other words, the loop and coupling expansions misalign, which is known as
Linde's infrared problem~\cite{Linde:1980ts}.
This problem can be bypassed by reorganising the perturbative expansion,
which is most economically achieved by treating
the zero mode separately in
a three-dimensional effective theory~\cite{Appelquist:1981vg,Nadkarni:1982kb}.
The transition from
fundamental four-dimensional, thermal QCD to
its three-dimensional effective theory EQCD is dubbed
{\em dimensional reduction}.
The continuum action of this effective theory reads%
\footnote{%
  The kinetic term commutators ensure the adjoint representation of the $\Phi$-fields.
  Henceforth, we will keep commutators implicit.
  In principle, a second quartic $\Phi$-operator
  $\lambda_2 \Tr \, \Phi^4$ is conceivable but linearly independent only for $\Nc > 3$.
  By focusing on QCD and in favour of numerical simplicity, we safely disregard the latter.
}
\begin{equation}
\label{eq:EQCD_cont_action}
S_{\rmii{EQCD},c} = \int_{\vec{x}}\bigg\{
    \frac{1}{2} \Tr\,F_{ij} F_{ij}
  + \Tr\,[D_i, \Phi] [D_i, \Phi]
  + \mDsq \Tr\,\Phi^2
  + \lambdaE^{ }(\Tr\,\Phi^2)^2
\bigg\}
  \, ,
\end{equation}
where
$\int_{\vec{x}} = \frac{1}{\gE^{4 \epsilon}} \int \! {\rm d}^d {\bf x}$,
$d=3-2\epsilon$ and
$D_i = \partial_i - i \gE A_i$
is the EQCD covariant derivative.
The former temporal component of the gauge field turns into
the scalar field {\em viz.} $\Phi = i A_0$ in the adjoint representation of SU(3) with
the acquired thermal mass $\mDsq$.

Since the resulting theory is static in three dimensions,
derivatives in temporal direction are absent.
Similarly, gauge invariance does not protect the $A^0$ or $\Phi$ field
from acquiring a mass.
The quartic coupling of $\Phi$ is a remnant of the four-$A^0$-interaction that arises at two-loop level in full-QCD perturbation theory.
This effective theory was originally proposed by Appelquist
and Pisarski~\cite{Appelquist:1981vg,Nadkarni:1982kb}.
Braaten and Nieto~\cite{Braaten:1995cm} presented a rigorous
perturbative procedure for constructing the effective theory
and determining its parameters $\mDsq$, $\lambdaE$, and $\gsq$ by matching; they also
named the theory `Electrostatic Quantum Chromodynamics' (EQCD).
The current matching is available at
$\rmO(g^6)$ for
$\gE^{2}$~\cite{Laine:2005ai},
$\mD^{2}$~\cite{Ghisoiu:2015uza} and
$\lambdaE^{ }$~\cite{Kajantie:1997tt},
with further improvements in~\cite{Laine:2018lgj,Laine:2019uua}.
The matching of these effective parameters and
the running coupling $g(\bmu / \Lambda_{\overline{\rmii{MS}}})^2$ are employed at $\OO(g^4)$,
while we use a renormalization scale of $\Lambda_{\overline{\rmii{MS}}} = 341$~MeV~\cite{Bruno:2017gxd},
obtained from $(2+1)$-flavor lattice simulations.

Since EQCD is a super-renormalizable theory, all amplitudes can be rendered finite by a finite number of counterterms. In our case, only the scalar mass $\mDsq$ receives a renormalization and therefore carries the only scale dependence of any parameter in the action. Using the dimensionful gauge coupling to set the scale $\overline\mu=\gsq$,
one can re-phrase all EFT parameters as dimensionless ratios
\begin{equation}
  x \equiv \frac{\lambdaE^{ }}{\gsq}
\;,\quad
y \equiv \frac{\mDsq}{\gfour} \Bigr|_{\bmu=\gsq}
\;.
\end{equation}

Concerning the matching of the operators in \eqref{int_gauge_corr},  dimensional reduction involves replacing
$F^{i0} \to i \left[ D^i,\Phi \right]$,
as mostly already outlined in~\cite{Moore:2020wvy}.
Furthermore, we replace the lightlike Wilson line $U_\rmii{A}$ with its
EQCD counterpart~\cite{CaronHuot:2008ni}
\begin{equation}
\label{eq:EQCD:mod:WL}
  \tilde{U}_\rmii{A}(L;0) = \mbox{P} \exp \Bigl( i\gE^{ }\int_0^L \! {\rm d} z\,
    \bigl(A_z^a(z) + i\Phi^a(z)\bigr) T_\rmii{A}^a
  \Bigr)
  \;,
\end{equation}
oriented along the $z$-axis and
suppressed constant transverse coordinates ($\xperp$).
With rotational invariance in the transverse plane, we find
\begin{align}
\label{Zg_EQCD}
\Zgtd &= -\frac{2T}{\dA} \int_0^\infty\!\! {\rm d} L\,L\,
  \Bigl\langle
    \Bigl(F_{xz}^a(L) + i (D_x \Phi(L))^a\Bigr)
    \,\tilde{U}^{ab}_\rmii{A}(L,0)\,
    \Bigl(F_{xz}^b(0) + i (D_x \Phi(0))^b\Bigr)
  \Bigr\rangle
\;\nn &=
  -\frac{4T}{\dA} \int_0^\infty\!\! {\rm d} L\,L\,
  \Bigl( -\EE + \BB + \EB \Bigr)
\;,
\end{align}
where
the explicit factor of $T$ accounts for the different
normalisation of the EQCD fields and the three different correlators are abbreviated as
\begin{align}
\label{def:cond:EE}
  \EE &\equiv \frac12
    \bigl\langle (D_x \Phi(L))^a\,\tilde{U}^{ab}_\rmii{A}(L,0)\,(D_x\Phi(0))^b \bigr\rangle
    \;,\\
\label{def:cond:BB}
  \BB &\equiv \frac12 \bigl\langle F_{xz}^a(L)\,\tilde{U}^{ab}_\rmii{A}(L,0) \, F_{xz}^b(0) \bigr\rangle
    \;,\\
\label{def:cond:EB}
  \EB &\equiv
      \frac{i}{2} \bigl\langle (D_x \Phi(L))^a\,\tilde{U}_\rmii{A}^{ab}(L,0) \, F_{xz}^b(0) \bigr\rangle
  \nn &
    + \frac{i}{2} \bigl\langle F_{xz}^a(L) \, \tilde{U}_\rmii{A}^{ab}(L,0) \, (D_x \Phi(0))^b \bigr\rangle
  \;.
\end{align}
An adjoint Wilson line between two operator insertions $\OO$ and $\OO'$ can be related to a pair of fundamental Wilson lines via
\begin{equation}
\label{eq:adj:fund}
  \langle \OO_a(x)\, U^{ab}_\rmii{A}(x;0)\, \OO'_b(0) \rangle =
  2 \, \Tr \, \langle
    \OO(x)\, U^{ }_\rmii{F}(x;0)\,
    \OO'(0)\, U^{-1}_\rmii{F}(x;0) \rangle
  \;.
\end{equation}
On the lattice it is more convenient to evaluate the correlators contributing to 
\eqref{Zg_EQCD} in the fundamental representation.
By absorbing a factor of $1/2$ from eq.~\eqref{Zg_EQCD}
into the correlators \eqref{def:cond:EE}--\eqref{def:cond:EB},
they can naturally be considered as living in the fundamental representation.

We will approach computing $\Zgtd$ from \eqref{Zg_EQCD} from two different sides:
perturbatively to NLO in EQCD or $\OO(\gsq)$, and
non-perturbatively in lattice EQCD.
At short separations $L$,
we expect the perturbative and non-perturbative lattice results to agree.
Treating short distances on the lattice is challenging.
The lattice
spacing must be kept several times smaller than the separation of interest
to avoid contamination from higher-dimension operators.
The lattice volume must be kept larger than a certain physical scale
to ensure that one maintains the right phase structure (the symmetric
phase of EQCD is only metastable, and the metastability is lost when
the volume gets too small \cite{Moore:2020wvy}).
And the precision
we demand becomes prohibitive, as the $\EE$ and $\BB$ correlators each
diverge as $1/L^3$, while the errors must be smaller than
of order $1/L^2$ for eq.~\eqref{Zg_EQCD} to converge.
At large separations, however, perturbation theory is supposed to become unreliable, and
data points directly obtained from lattice simulations or fits of
large-$\gsql$ models can provide further insight.

%
\section{Perturbative determination of $\Zgtd$ at NLO in EQCD}
\label{sec:matching}

\begin{figure}[t]
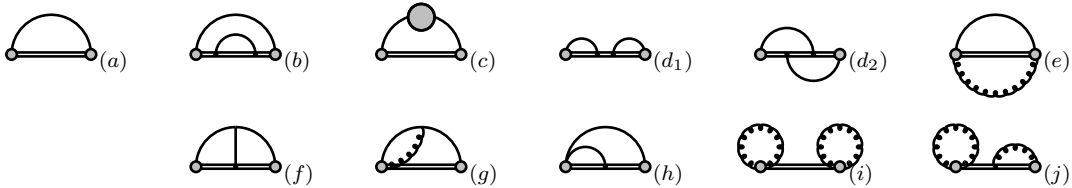

  \centering
  \begin{align*}
    \TopoWR(\Lsri,\Asai)_{(a)}\;
    && &
    \ToptWBB(\Lsri,\Asai,\Asai)_{(b)}\;
    &&
    \ToptWSBB(\Lsri,\Asai,\Asai,\Asai,\Asai)_{(c)}\;
    &&
    \ToptWSBSB(\Lsri,\Asai,\Asai)_{(d_1)}\;
    &&
    \ToptWSBBS(\Lsri,\Asai,\Asai)_{(d_2)}\;
    &&
    \ToptWS(\Lsri,\Asai,\Agliii)_{(e)}\;
    \nn[3mm]
    && &
    \ToptWM(\Lsri,\Asai,\Asai,\Lsai)_{(f)}\;
    &&
    \ToptWSal(\Lsri,\Asai,\Asai,\Agliii)_{(g)}\;
    &&
    \ToptWal(\Lsri,\Asai,\Asai)_{(h)}\;
    &&
    \ToptWTT(\Lsri,\Agliii,\Agliii)_{(i)}\;
    &&
    \ToptWTR(\Lsri,\Agliii,\Agliii)_{(j)}\;
  \end{align*}
  \caption{%
    Diagrams contributing to leading and next-to-leading order to
    the EQCD force-force correlator $\Zgtd$~\eqref{Zg_EQCD}.
    An external gray shaded vertex denotes a
    $D^{x}\Phi$ or
    $F^{xz}$ insertion;
    internal 2-point blobs the respective self-energy;
    a double line a Wilson line; and
    a curly line a spatial gauge boson ($A_i$). 
    A solid line is a placeholder for either 
    an adjoint scalar ($\Phi$) or
    a spatial gauge boson (otherwise curly).
  }
  \label{fig:diagrams}
\end{figure}

As motivated above, our task is to
match the gluonic EQCD correlator~\eqref{Zg_EQCD} to its
full QCD counterpart.
In fact, the UV contributes to $\Zg$
since the latter contains an integral over length scales in the domain
$L \in \left[0,\infty \right)$.
The ultraviolet (UV) region is precisely where we expect corrections from full QCD that are
not included in EQCD.
Fortunately, these contributions should be under better perturbative control than
the IR regime due to asymptotic freedom.
The corresponding diagrams are compiled in fig.~\ref{fig:diagrams}.

Computing the three correlators in EQCD,
$\EE$, $\EB$, and $\BB$, is only possible at finite values of $\gsql$ on the lattice.
Beyond the feasible range of $\gsql$ on the lattice, one has to rely on models.
For large $\gsql$, one can fit the largest-$\gsql$ lattice data points to asymptotic models, as done in~\cite{Moore:2020wvy} and updated in sec.~\ref{sec:latt}.
This regime produces a small contribution because
the correlators decay exponentially here.
For small $\gsql$, perturbation theory is supposed to work in EQCD. Since the three-dimensional coupling $\gsq$ carries mass dimension one, and the correlators carry mass dimension three, dimensional analysis tells us that the tree-level EQCD expressions can go as $1/L^3$ in the small-$\gsql$-limit at worst, whereas the one-loop level can contain $\gsq/L^2$ at worst, and all higher loop levels are $\OO(\gfour/L)$ at worst.
The $L\,{\rm d} L$ integration leading to $\Zgtd$ in \eqref{Zg_EQCD} can therefore receive UV-divergent contributions only from the $\OO(1/L^3)$ LO terms or from the $\OO(\gsq/L^2)$ NLO terms.%
\footnote{
  When we talk about leading or next-to-leading order in the following,
  we refer to orders in the EQCD perturbative expansion in $\gsq$, being related but
  not to be confused with the full QCD perturbative expansion in $g$.
  We will thus drop the previously adapted -in-EQCD specifiers.
}
All higher-order contributions to $\Zgtd$ are short-distance finite.
Therefore, a one-loop analytical calculation of the three correlators is not only required
quantitatively for increasing the agreement of lattice data and perturbation theory at small $\gsql$, but also
qualitatively for a comprehensive treatment of all possible UV divergences.
Note that this short-distance region is where EQCD no longer
provides a good description of full QCD.
Nevertheless, an accurate
treatment of this region will be needed when we carry out the matching
to the full four-dimensional theory, which we leave to a future publication.

Below, we present the next-to-leading order perturbative calculation in EQCD.
Its mass parameter $\mDsq$ is fully resummed and {\em not} treated as a perturbation of $\OO(g^2)$, while also
fully taking the quartic $\Phi$-vertex into account.
Furthermore, we employ momentum-space
gauge and
adjoint scalar propagators
\begin{align}
\label{eq:gl:prop:mom}
\langle A^a_i(\bop) A^b_j(\bq) \rangle &=
  \frac{\delta^{ab}(2\pi)^3\delta^{(3)}(\bop+\bq)}{p^2}\Big(
    \delta_{ij} - (1 - \xi)\frac{p_{i}p_{j}}{p^2}
  \Big)
  \;, \\
\label{eq:sc:prop:mom}
\langle \Phi^a(\bop) \Phi^b(\bq) \rangle &=
  \frac{\delta^{ab}(2\pi)^3\delta^{(3)}(\bop+\bq)}{p^2 + \mD^2}
\;,
\end{align}
where the former is in general covariant gauge with gauge fixing parameter $\xi$ and 
$\vec{p}$ is a three-dimensional vector with modulus $|\vec{p}| = 0$.
We will generally use
Feynman gauge $\xi = 1$, and this should be assumed except in
expressions with explicit $\xi$ dependence.

%
\subsection{The leading order}
\label{sec:LO}

The leading $\OO(\gE^0)$ contribution is also the free solution to $\Zgtd$,
consists of the 
graphs from fig.~\ref{fig:diagrams}$(a)$, and
was calculated in~\cite{Moore:2020wvy}.

After inserting the propagators from
eqs.~\eqref{eq:gl:prop:mom} and \eqref{eq:sc:prop:mom},
and explicitly separating integration momenta as
$\vec{p} = \vec{p}_{\perp}^{ } + p_{z}^{ }\vec{e}_{z}^{ }$,
the remaining integrals are Fourier transforms of the form
\begin{equation}
\label{eq:LO:int}
  \int_{\vec{p}} \frac{e^{iLp_{z}}}{[p^2 + \mD^2]^{\alpha}}\bigl(p^2 - p_{z}^2\bigr)
  \;,\quad
  \int_{\vec{p}} \frac{e^{iLp_{z}}}{[p^2]^{\alpha}}\bigl(p^2 + (d-2) p_{z}^2\bigr)
  \;,
\end{equation}
where the gauge dependence of $\BB$ cancels due to
the antisymmetry of the field strength tensor.
The integrals over $p_{z}^{ }$ and $p_\perp$ can be evaluated with the residue theorem
and in dimensional regularization respectively with $d=3-2\epsilon$, 
and results are collected in appendix~\ref{sec:masters}.
Contracting color indices $\delta^{aa} = 2\CA\CF$ yields
for the diagonal correlators
\begin{align}
\label{EE:free}
\TopoWR(\Lsri,\Asa3) =
2\times (a)^{\rmii{EE}} &=
  \partial_{x}\partial_{x'}
  \Tr\bigl\langle
     \Phi^a(x,L)
     \Phi^a(x',0)
  \bigr\rangle\bigr|_{x,x'\to 0}
  \nn &=
  \frac{2\CA\CF}{4\pi L^3} \e^{- \mD^{ }L} \left( 1 + \mD  L \right)
  \;, \\
\label{BB:free}
\TopoWR(\Lsri,\Agl3) =
2\times (a)^{\rmii{BB}} &=
  \partial_{x}\partial_{x'}
    \Tr\bigl\langle
      A_z^a(x,L)
      A_z^a(x',0)
    \bigr\rangle \bigr|_{x,x'\to 0}
    + \partial_{z}\partial_{z'}
    \Tr\bigl\langle
      A_x(0,z)
      A_x(0,z')
    \bigr\rangle\bigr|_{\substack{z=L\\z'\to 0}}
  \nn &=
  - \frac{2\CA\CF}{4\pi L^3}
  \;,
\end{align}
with Wilson line endpoints
$\vec{x} = (\vec{0}_\perp,L)$ and $\vec{x}'=(\vec{0}_\perp,0)$.
Here and in the following we draw and evaluate
each diagram (e.g.~(a)) 
in its adjoint form which is, as we argued,
twice the fundamental form in the definitions of 
$\langle EE\rangle$,
$\langle BB\rangle$, and 
$i\langle EB\rangle$ below eq.~\eqref{def:cond:EE}.
This simplifies the graphical notation and color evaluation. 
In turn,
we account for the factor of 2 explicitly in
contributions of adjoint diagrams to any channel
e.g.\
$\EE$, as $2 \times (a)^{\rmii{EE}}$.
Henceforth,
a curly line denotes a spatial gauge boson ($A_i$) and
a solid line an adjoint scalar ($\Phi$).

The $\mathbb{Z}_2$ symmetry of EQCD requires the number of $\Phi$ fields to be even.%
\footnote{
  The $\mathbb{Z}_2$ symmetry of EQCD arises 
  from its action, eq.~\eqref{eq:EQCD_cont_action}, being an 
  even function of $\Phi$. It is not to be confused with
  the $\mathbb{Z}_3$ center symmetry of  full QCD, which is explicitly broken in EQCD~\cite{Vuorinen:2006nz}.
}
Therefore,
at leading order the mixed $\EB$ correlator vanishes and
at next-to leading order only graphs with a single $\Phi$ sourced from the Wilson line contribute.
Similarly, only graphs with zero or two $\Phi$ fields
sourced from the Wilson line contribute at NLO to
the $\EE$ and $\BB$ operators (cf.~fig.~\ref{fig:diagrams}).

The physical short distance behavior of
the (four-dimensional) gauge condensate $\Zg$ is perturbative,
as detailed in sec.~\ref{sec:power:counting}.
Therefore the UV of $\Zg$ is described in four-dimensional thermal QCD, as
it is dominated by energy scales of order $T$ rather than $\mD$. 
When it comes to an IR-consistent treatment of longer-range corrections,
EQCD comes into play. 
Since both lattice and perturbative EQCD results for $\Zgtd$
contain an (unphysical) UV limit, while conversely
the $T$-scale contributions
contain an incorrect IR limit, as they integrate
over the IR momentum region without resummation (see eq.~\eqref{zglo}),
a subtraction scheme is needed to avoid double counting. 

The UV limit of the EQCD correlators~\eqref{Zg_EQCD},
can be extracted from the small $\mD L$ asymptote of the perturbative contributions.
At leading order starting from \eqref{EE:free} and \eqref{BB:free}
and taking
$\langle\dots\rangle^{\rmii{subtr}} = \lim^{ }_{\mD L\to 0}\langle\dots\rangle$
this gives rise to the subtractions
\begin{equation}
\label{eq:LO:UV}
    \EE_\rmii{LO}^\rmii{subtr} =
  - \BB_\rmii{LO}^\rmii{subtr} =
  \frac{\CA\CF}{4\pi L^3}
  \;,
\end{equation}
which contain all non-resummed (i.e.\ $\mD \to 0$)
EQCD contributions at this order and
should be removed from $\Zgtd$.
These subtractions also eliminate the UV divergences in the EQCD result, 
which are not considered to be physical: they correspond to the 
linear-in-$\mu_h$ term in eq.~\eqref{nloline} 
(see footnote~\eqref{foot_lo_div}) and
are already included in the unresummed zero-mode contribution to the LO-in-QCD
contribution to $\Zg$.

%
\subsection{Relation to momentum broadening kernel}
\label{sec:relation:qhat}
We now comment on the relation
between $\Zg$ and the transverse momentum broadening coefficient $\hat{q}$.
The latter is defined  as 
the second moment of the broadening probability $P(q_\perp)$ 
or of the transverse
scattering kernel $\mathcal{C}(q_\perp)$ up to some
UV cutoff $q_\mathrm{max}$.
Both can be derived
from a Wilson loop in the $(x^+,\vec{x}_\perp)$ plane~\cite{%
  CasalderreySolana:2007qw,CaronHuot:2008ni,DEramo:2010wup,Benzke:2012sz}.
Upon writing $q_\perp^2$ as transverse derivatives acting on this Wilson loop,
one obtains for a fundamental source~\cite{Benzke:2012sz} 
\begin{align}
  \hat{q} =& \;
    \frac{g^2}{\Nc}
    \int \! \frac{{\rm d}^2q_\perp}{(2\pi)^2}
    \int \! {\rm d}^2 x_\perp
    \int_{-\infty}^\infty \! {\rm d} x^+ e^{i\vec{q}_\perp\cdot \vec{x}_\perp}
  \Bigl\langle\mathrm{Tr}\;
    \bar{U}_\rmii{F}^{ }(0;\vec{x}_\perp)^{ }_{-\infty}
    U_\rmii{F}^{ }(-\infty;x^+)^{ }_{\vec{x}_\perp}
    v_\mu^{ } F^{\mu\nu}_a (x^+,\vec{x}_\perp)
  \nn &\hspace{20pt} \times
    U_\rmii{F}^{ }(x^+;+\infty)^{ }_{\vec{x}_\perp}\,
    \bar{U}_\rmii{F}^{ }(\vec{ x}_\perp;0)^{ }_{+\infty}
    U_\rmii{F}^{ }(+\infty,x^+)^{ }_{0}
    v_\rho^{ } F^\rho_{b\,\nu}(0,0) U_\rmii{F}(0;-\infty)^{ }_{0}
  \Bigr\rangle
  \;,
  \label{int_gauge_corr_qhat_loop}
\end{align}
where
$U_\rmii{F}^{ }(b^+;a^+)_{\vec{y}_\perp}$
is a fundamental Wilson line in the ${}^+$ direction
from $a^+$ to $b^+$ at fixed $\vec{y}_\perp$ coordinate, while
the ${}^-$ coordinate is fixed to the same value for all fields in the expression above. 
$\bar{U}_\rmii{F}(\vec{b}_\perp;\vec{a}_\perp)_{y^+}$ is instead a fundamental Wilson line in the $\perp$ direction
from $\vec{a}_\perp$ to $\vec{b}_\perp$ at fixed $y^+$ coordinate.
If the $\vec{q}_\perp$ integration is taken with infinite cutoff, i.e.\ in 
dimensional regularization, the resulting $\delta^{(2)}(\boldsymbol x_\perp)$ will squeeze
the Wilson loop in eq.~\eqref{int_gauge_corr_qhat_loop} into a Wilson line,
{\em viz.}~\cite{Benzke:2012sz}
\begin{equation}
\label{int_gauge_corr_qhat}
  \hat{q} = \frac{g^2}{\Nc}
  \int_{-\infty}^\infty \! {\rm d} x^+
  \Bigl\langle\mathrm{Tr}\;
    U_\rmii{F}^{ }(-\infty;x^+)
    v_\mu^{ } F^{\mu \nu}_a (x^+)\,U_\rmii{F}^{ }(x^+;0)\,
    v_\rho^{ } F^\rho_{b\,\nu}(0)\,U_\rmii{F}^{ }(0;-\infty)
  \Bigr\rangle
  \;,
\end{equation}
where we dropped the -- now fixed -- transverse coordinate of the Wilson lines.
We can now apply dimensional reduction on eq.~\eqref{int_gauge_corr_qhat} to find
the soft, three-dimensional contribution to $\hat{q}$
\begin{align}	
\label{qhat_EQCD} 
  \hat{q}^{\rmii{3d}} =
    \frac{2\gsq\CF^{ }}{\dA^{ }}
    \int_{-\infty}^\infty \! {\rm d} L\,\biggl\{&
      - \Bigl\langle (D_x\Phi(L))^a \tilde{U}^{ab}_A(L;0)(D_x\Phi(0))^b\Bigr\rangle
      + \Bigl\langle F_{xz}^a(L) \tilde{U}^{ab}_A(L;0)F_{xz}^b(0) \Bigr\rangle
    \nn & 
    +i \Bigl\langle (D_x\Phi(L))^a \tilde{U}^{ab}_A(L;0)F_{xz}^b(0)\Bigr\rangle
    \nn & 
    +i \Bigl\langle F_{xz}^a(L) \tilde{U}^{ab}_A(L;0)(D_x\Phi(0))^b\Bigr\rangle
    \biggr\}
    \;.
\end{align}
This shows that $\hat{q}$ is related to $\Zgtd$ by a change in the
overall prefactor and in the integration, which is
$\int_{-\infty}^{+\infty}{\rm d}L$ rather than
$\int_{0}^{+\infty}{\rm d}L\,L$.

We stress that eq.~\eqref{qhat_EQCD} is to be understood
in dimensional regularization only. If this regularization is not taken, divergences
arise at $L=0$ in the ${\rm d}L$ integration. This is apparent at LO
from eqs.~\eqref{EE:free} and \eqref{BB:free}, which would
diverge as ${\rm d}L/L^3$ at the origin.
We have however checked that,
if eqs.~\eqref{EE:free} and \eqref{BB:free} are evaluated in $d$
dimensions and then plugged in eq.~\eqref{qhat_EQCD} and integrated in ${\rm d}L$, 
they yield
\begin{equation}
\label{loqhatsoft}
\hat{q}^{\rmii{3d}}_\rmii{LO} =
  \frac{\gsq\CF^{ }\mD^2}{4\pi}\Bigl(
    \frac{1}{\epsilon}
  + \ln\frac{\bmu^2}{\mD^2}
  \Bigr)\;,
\end{equation}
where our conventions for dimensional regularization are summarized in 
appendix~\ref{sec:masters}.
This agrees with a dimensionally-regularized
evaluation of the LO soft contribution to $\hat{q}$~\cite{Aurenche:2002wq,CaronHuot:2008ni},
i.e.
\begin{equation}
\label{loqhatsoftmomspace}
\hat{q}^{\rmii{3d}}_\rmii{LO} =
  \gsq\CF^{ }
  \int_{\vec{q}_\perp}
  q_\perp^2\Bigl(
      \frac{1}{q_\perp^2}
    - \frac{1}{q_\perp^2+\mD^2}
  \Bigr)\;,
\end{equation}
where 
$
\int_{\vec{q}_{\perp}}\equiv
\mu^{3-d}\int\frac{{\rm d}^{d-1} \vec{q}_{\perp}^{ }}{(2\pi)^{d-1}}
$.
This evaluation highlights another aspect that will become relevant
when performing the same check for our NLO evaluation of $\Zgtd$,
namely that squeezing the Wilson loop onto itself through dimensional
regularization obfuscates some cancellations that would be 
apparent had the integrals been performed in a different order.
With that we mean that the UV divergence of 
$\hat{q}^{\rmii{3d}}_\rmii{LO}$ is logarithmic,
as it is well known and clear from eq.~\eqref{loqhatsoftmomspace}.
However, inserting eqs.~\eqref{EE:free} and \eqref{BB:free} in eq.~\eqref{qhat_EQCD}
leads to severe ${\rm d} L/L^3$ UV divergences: the cancellation of the leading
UV behavior between the magnetic and electric contributions --
made explicit by the two terms in brackets in eq.~\eqref{loqhatsoftmomspace} --
seems lost.
However, let us return to eq.~\eqref{BB:free}:
had we taken the ${\rm d} L$ integration before taking
the ${\rm d}^3{\bf p}$ integrals in eq.~\eqref{eq:LO:int},
the $\partial_{z}^{ }\partial_{z'}$ part would simply vanish due to the $\delta(p_z)$
resulting from the ${\rm d} L$ integration, 
thereby making apparent the cancellation between the leading UV behavior of
$\EE$ and $\BB$. 

%
\subsection{The next-to-leading order}
\label{sec:NLO}

At next-to-leading-order (NLO) multiple graphs contribute to $\Zgtd$ in EQCD as
depicted in fig.~\ref{fig:diagrams}$(b)$--$(j)$.
Relegating their explicit evaluation to appendix~\ref{sec:NLO:details},
here we merely sum their individual contributions which yields
for the $\EE$-correlator%
\footnote{\label{footnote_eamonn}%
We thank Eamonn Weitz
for spotting a missing factor of $\mD L$ in the $\lambdaE$-proportional
term of this result and of eqs.~\eqref{eq:c:EE} and \eqref{eq:c:EE:UV} in an earlier version of this
paper. The error affected also eq.~\eqref{eq:NLO:UV} and
the last digits of the numerical results in sec.~\ref{sec:latt}.
}
\begin{align}
\label{PT_EE_final}
\frac{(4\pi L)^2 \mD L}{\gsq\CA^{2}\CF^{ }}\times
\EE_\rmii{NLO} &=
    \e^{-2\mD L} (1 + \mD L)
  + \e^{\mD L} (1 - \mD L) E_{1}(2\mD L)
  \nn &
  - \e^{-\mD L} \Bigl[ 1 - \gammaE - (\mD L)^2 \Bigl(\frac{3}{2} - \gammaE\Bigr)\Bigr]
  \nn[2mm] &
  - (\mD L)\,\e^{-\mD L} \bigl[
      (1 + \mD L) E_{1}(\mD L)
    - \ln{2}
    + \mD L \ln(\mD L)
  \bigr]
  \nn[2mm] &
  + \e^{-\mD L} \ln(2\mD L)
  -x\, (\mD L)^2\,
  \e^{-\mD L}\frac{\CA - 6 \CF}{48\CA}
  \;,
\end{align}
where
$E_{1}$ is the exponential integral function~\eqref{eq:E1} and
$x=\lambdaE/\gE^2$ the dimensionless scalar self-coupling.
For the $\BB$-correlator, we find
\begin{align}
\label{PT_BB_final}
\frac{(4\pi L)^2 \mD L}{\gsq\CA^{2}\CF^{ }} \times
\BB_\rmii{NLO} &=
  - \e^{-\mD L}
  - \frac{1}{12}\e^{-2\mD L}\bigl(1 + 2(\mD L) - 4(\mD L)^2 \bigr)
  + \frac{13}{12}
  \nn &
  - \mD L \bigl(1 - \gammaE - \ln (\mD L) - E_{1}(\mD L)\bigr)
  - \frac{2}{3} (\mD L)^3 E_{1}(2 \mD L)
  \;.
\end{align}
Finally, our perturbative prediction for $\EB$ reads
\begin{align}
\label{PT_EBBE_final}
  \EB_\rmii{LO} &= 0
  \;,\\
\frac{(4\pi L)^2}{\gsq\CA^{2}\CF^{ }} \times
\EB_\rmii{NLO} &=  - 2 \e^{- \mD L} \mbox{Shi}(\mD L)
  \;,
\end{align}
where the hyperbolic sine integral function is defined as
$\mbox{Shi}(x) = \int_0^x \! {\rm d} t \frac{\sinh (t)}{t}$.

Both the LO (see eqs.~\eqref{EE:free} and \eqref{BB:free}) and NLO correlators have undergone several crosschecks.
One immediate crosscheck is gauge independence.
While individual contributions are indeed explicitly gauge dependent,
in the overall summation of the integrand of $\Zgtd$
the gauge fixing parameter $\xi$ of eq.~\eqref{eq:gl:prop:mom} duly cancels
at the integral level.

Another possible source of error is the regularization of divergences.
To this end, we extracted UV and IR divergences separately
by computing diagrams both in position and momentum space.
Where possible we verified individual results by comparison to literature in
the soft limit $\frac{p^2}{\mD^2}\ll 1$,
for instance to~\cite{Giovannangeli:2005rz} for diagram~$(c)$.
 
In a final non-trivial crosscheck, we recovered
the NLO result for the jet quenching parameter $\hat{q}$ from~\cite{CaronHuot:2008ni}
by relying on its relation with $\Zgtd$;
see sec.~\ref{sec:relation:qhat}.
As we discuss there, that relation holds in principle in dimensional regularization only.
However, we have not performed the entire NLO determination of $\Zgtd$
in an arbitrary number of dimensions, so we cannot simply take the final
${\rm d}L$ integral as in eq.~\eqref{qhat_EQCD}.
We have instead worked at the integral level:
as the previous subsection discusses, these integrals simplify greatly if
the ${\rm d}L$ integral is taken before the ${\rm d}^3\vec{p}$ ones.
Finally, we have been able to exploit the
fact that the NLO soft contribution to $\hat{q}$, due to the super-renormalizable nature
of EQCD, is free of logarithmic divergences.
It only contains a linear divergence
and is finite in dimensional regularization, so that we have been able
to carry out the remaining ${\rm d}^2 \vec{p}_\perp$
integrals and recover the result of~\cite{CaronHuot:2008ni}.

The determination of the UV limit of the EQCD correlators at next-to-leading order
changes qualitatively compared to leading order~\eqref{eq:LO:UV}.
The correlator $\EB$ is non-vanishing,
and the expressions become more complicated.
While the scalar self-coupling $\lambdaE$ enters eq.~\eqref{PT_EE_final} through the scalar self-energy,
it fortunately does not contribute any $\sim 1/L^2$ terms that would need to be subtracted.

The asymptotic small-$\mD L$ limit of the $\EE$
\begin{equation}
\label{eq:NLO:UV}
  \EE_\rmii{NLO}^\rmii{subtr} =
    2\frac{\gsq\CA^{2}\CF^{ }}{(4\pi L)^2}
  \;,
\end{equation}
indeed diverges in the UV ($L\to 0$) and
can therefore be subtracted from the EQCD result for $\Zgtd$.
Correlators $\BB$ and $\EB$ are at worst $\OO(L^{-1})$ in the UV at
NLO in EQCD and therefore do not contribute any
divergences to $\Zgtd$ and thus require no subtraction.
The two correlators containing $E$-field insertions give an
IR-finite contribution to $\Zgtd$, since they are not directly connected to the
(inherently non-perturbative) magnetic sector of QCD at 
$\OO(g^2)$ of the perturbative expansion, while $\BB$ has 
such direct connection at this order: it behaves
as $(\ln(L)+c)/L^2$ for large $L$ and gives
thus rise to an IR-divergent contribution to $\Zgtd$. 
Lattice EQCD data does 
not suffer from IR divergences, therefore we are not impacted 
by perturbative IR divergences as long as we switch from
perturbation theory to the lattice at sufficiently small $\gsql$.
However, the necessary subtraction $\EE_\rmii{NLO}^\rmii{subtr}$ in 
\eqref{eq:NLO:UV} additionally introduces, if integrated to infinite 
$L$, an IR divergence which will remain necessary for a lattice EQCD 
treatment due to UV divergences of the lattice data.
In the definition of $\Zgtd$~\eqref{Zg_EQCD} the final integration over ${\rm d}L\,L$
of $\OO(L^{-2})$ terms delivers a logarithmic IR divergence
which we expect to cancel in the matching with a corresponding IR divergence in full QCD.
Consequently, $\Zgtd$, the result within EQCD presented in this publication, will still depend on an IR cutoff which we will introduce in the next section, even though the dependence on the regulator should vanish in the final, fully matched result for $\Zg$.

%
\section{EQCD lattice results}
\label{sec:latt}
In this section,
we extend the results of~\cite{Moore:2020wvy} to 
include $\EB$ and then compare the contiuum-extrapolated
data to our perturbative determination.
We then provide a UV-subtracted,
scheme-dependent number for the IR contribution to $\Zg$;
the scheme dependence
will disappear once the NLO $T$-scale contribution will become available.
For the details of the used implementation of lattice 
EQCD we ask the reader to consult appendix~\ref{sec:sim:details} and
the references therein.

%
\subsection{Lattice determination and continuum extrapolation}
Having identified $\EB$ as an additional non-trivial, but numerically subdominant correlation function contributing to $\Zg$, there is a need to measure $\EB$ from the lattice.
Just as in the continuum case, non-trivial operator mixing also occurs beyond leading order in lattice perturbation theory.
The operator-mixing of the three different continuum operators
$\langle\dots\rangle_\rmi{c}$ into a single lattice operator
$\langle\dots\rangle_\rmii{L}$ follows from
\begin{align}
  \label{latt_op_mix}
\left[ \begin{array}{c}
\hp{i}\EE_\rmii{L} \\
\EB_\rmii{L} \\
\hp{i}\BB_\rmii{L} \\
\end{array} \right]
&= \; \e^{\ZP^{ } \gfour L a}
\left[ \begin{array}{ccc}
    1 + 2\gsqa \ZEE^{ } & \gsqa \ZEB^{ } & 0 \\
    \gsqa \ZBE^{ } & 1 + \gsqa (\ZEE^{ } + \ZBB^{ })  & \gsqa \ZEB^{ } \\
    0 & \gsqa \ZBE^{ } & 1 + 2\gsqa \ZBB^{ } \\
\end{array}
\right]
\left[ \begin{array}{c}
\hp{i}\EE_\rmi{c} \\
\EB_\rmi{c} \\
\hp{i}\BB_\rmi{c} \\
\end{array} \right]
\nn[2mm] &
+ \OO(\gfour a^2)
\, ,
\end{align}
with
four multiplicative renormalization constants
$\ZEE$, $\ZEB$, $\ZBE$, $\ZBB$, and
the perimeter-law contribution of the modified Wilson line $\ZP$.
It is necessary to have four constants since the coefficient associated with turning an $E$-field into a $B$-field does not have to be the same as vice versa.
Adapting the fitting procedure outlined in~\cite{Moore:2020wvy} to \eqref{latt_op_mix},
we can repeat the grand fit and end up with a decent fit likelihood corresponding to $\chi^2 / \mathrm{d.o.f.} = 400 / 247 \approx 1.6$.
The continuum results are
displayed in tab.~\ref{tab:sim_res}.%
\footnote{
  To avoid confusion, we updated the values of tab.~2 in the arXiv-version of~\cite{Moore:2020wvy},
  except from the $\EB$ measurement and the $\gsql=0.25$ data point of $\BB$.
  Note also that this publication uses slightly different conventions for the normalization of the fields, such that $\Zgtd$ has now mass dimension 1 instead of 2 in \cite{Moore:2020wvy}.
}
Beyond the operator mixing that also involves the $\EB$-contribution, the convergence of the $\BB$-operator to the continuum was accelerated compared to \cite{Moore:2020wvy} by replacing the clover-clover $\BB$-correlator with a single-plaquette expectation value. 
Although this decreases statistical power (4 instead of 16 combinations of single plaquettes), it renders all four measured correlators at the correct lattice spacing, there are no more contaminations of operators at the wrong separation. With this in hand, we were able to also achieve a valid continuum limit for $\BB$ at $\gsql=0.25$.
The discussion of different sources for errors in \cite{Moore:2020wvy} also applies here.
\begin{table}[ptbh]
\centering {\small
\begin{tabular}{|c|c|c|c|c|c|c|}
  \hline
  &
  \multicolumn{3}{c|}{$T=250~\mathrm{MeV},\; \Nf=3$}
  &
  \multicolumn{3}{c|}{$T=500~\mathrm{MeV},\; \Nf=3$}
  \\
  &
  \multicolumn{3}{c|}{$x=0.08896,\; y=0.452423$}
  &
  \multicolumn{3}{c|}{$x=0.0677528,\; y=0.586204$}
  \\
  \hline
  $\gsql$ & $\EE$ & $\BB$ & $\EB$ & $\EE$ & $\BB$ & $\EB$ \\
  \hline
  0.25 &
  $22.761(41)$ & $-20.136(47)$ & -- &
  $22.559(35)$ & $-20.107(53)$ & -- \\
  0.5 &
  $3.1275(60)$ & $-2.5672(98)$ & $ 0.1202(60)$ &
  $3.0230(61)$ & $-2.519(10)$ & $  0.0255(50)$ \\
  0.75 &
  $1.0116(32)$ & $-0.756(11)$ & $  0.0458(52)$ &
  $0.9436(30)$ & $-0.7349(91)$ & $-0.0042(37)$ \\
  1.0 &
  $0.4539(25)$ & $-0.3217(33)$ & $ 0.0187(30)$ &
  $0.4092(19)$ & $-0.3057(33)$ & $-0.0115(23)$ \\
  1.5 &
  $0.1470(19)$ & $-0.0995(51)$ & $ 0.0013(27)$ &
  $0.1186(20)$ & $-0.0822(40)$ & $-0.0107(28)$ \\
  2.0 &
  $0.0713(31)$ & $-0.0255(72)$ & $ 0.0066(34)$ &
  $0.0448(19)$ & $-0.0217(56)$ & $-0.0046(46)$ \\
  2.5 &
  $0.033(13)$ & $0.005(42)$ & $-0.026(20)$ &
  $0.020(11)$ & $0.027(30)$ & $-0.013(19)$ \\
  3.0 &
  $0.036(20)$ & $0.034(46)$ & $ 0.076(39)$ &
  $0.023(17)$ & $0.060(56)$ & $-0.002(27)$ \\
  \hline
  $\!\vphantom{\bigg|} \Bigl( \frac{\Zgtd}{\gsq} \Bigr)_\rmii{cutoff}\!$ &
  \multicolumn{3}{c|}{$0.085(37)(12)(2)$} &
  \multicolumn{3}{c|}{$-0.023(36)(14)(1)$}
  \\
  \hline
  $\!\!\vphantom{\bigg|} \Bigl( \frac{\Zgtd}{\gsq} \Bigr)_\rmii{full}\!\!$ &
  \multicolumn{3}{c|}{$0.085(37)(12)(2)$} &
  \multicolumn{3}{c|}{$-0.024(36)(14)(1)$}
  \\
  \hline
  \hline
  &
  \multicolumn{3}{c|}{$T=1~\mathrm{GeV},\; \Nf=4$} &
  \multicolumn{3}{c|}{$T=100~\mathrm{GeV},\; \Nf=5$}
  \\ &
  \multicolumn{3}{c|}{$x=0.0463597,\; y=0.823449$}&
  \multicolumn{3}{c|}{$x=0.0178626,\; y=1.64668$}
  \\
  \hline
  $\gsql$ & $\EE$ & $\BB$ & $\EB$ & $\EE$ & $\BB$ & $\EB$ \\
  \hline
  0.25 &
  $22.290(33)$ & $-19.989(53)$ & -- &
  $21.588(34)$ & $-19.849(51)$ & -- \\
  0.5 &
  $2.8991(48)$ & $-2.4833(88)$ & $-0.0487(43)$ &
  $2.6111(59)$ & $-2.4214(88)$ & $-0.1474(46)$ \\
  0.75 &
  $0.8686(22)$ & $-0.7150(69)$ & $-0.0431(27)$ &
  $0.7107(35)$ & $-0.673(10)$ & $ -0.0813(40)$ \\
  1.0 &
  $0.3574(18)$ & $-0.2922(30)$ & $-0.0345(15)$ &
  $0.2616(20)$ & $-0.2750(30)$ & $-0.0534(20)$ \\
  1.5 &
  $0.0957(15)$ & $-0.0831(50)$ & $-0.0205(24)$ &
  $0.0530(12)$ & $-0.0580(32)$ & $-0.0179(16)$ \\
  2.0 &
  $0.0317(21)$ & $-0.02885(59)$ & $-0.0048(33)$ &
  $0.0133(16)$ & $-0.0232(38)$ & $ -0.0072(19)$ \\
  2.5 &
  $ 0.0179(93)$ & $-0.033(33)$ & $-0.011(17)$ &
  $-0.0020(59)$ & $-0.006(25)$ & $-0.0119(84)$ \\
  3.0 &
  $ 0.0223(13)$ & $-0.010(35)$ & $ -0.008(16)$ &
  $-0.0012(64)$ & $-0.024(20)$ & $-0.0157(92)$ \\
  \hline
  $\!\vphantom{\bigg|} \Bigl( \frac{\Zgtd}{\gsq} \Bigr)_\rmii{cutoff}\!$ &
  \multicolumn{3}{c|}{$-0.021(32)(4)(3)$} &
  \multicolumn{3}{c|}{$-0.162(15)(5)(2)$}
  \\
  \hline
  $\!\!\vphantom{\bigg|} \Bigl( \frac{\Zgtd}{\gsq} \Bigr)_\rmii{full}\!\!$ &
  \multicolumn{3}{c|}{$-0.026(32)(4)(3)$} &
  \multicolumn{3}{c|}{$-0.179(15)(5)(2)$}
  \\
  \hline
\end{tabular}
}
\caption{%
  Continuum-extrapolated results for the correlators
  $\EE/\gsix$,
  $\BB/\gsix$, and
  $\EB/\gsix$, at four tuples $(T,\Nf)$ of
  temperature and
  number of massless quark flavors over a range of separations $\gsql$.
  $(\Zgtd)_\rmii{latt}$ sums the integrated correlators according to \eqref{Zg_EQCD}
  with small-$\gsql$ cutoff for $\EE$ and $\BB$ at $\gsqlmin^{ }=0.25$ and
  for $\EB$ at $\gsqlmin^{ }=0.5$.
  Note that the scalar field $\Phi$ on the lattice
  is rescaled such that it is dimensionless. Therefore, the three correlators on 
  the lattice -- and $\Zgtd{}$ accordingly -- feature slightly different
  normalizations than in perturbation theory
  (cf.~\eqref{Zg_EQCD}, app.~\ref{sec:sim:details}).
  $(\Zgtd)_\rmii{cutoff}$ contains the entire integration of lattice data points, LO subtraction and long-$\gsql$-tail integration in the range $\gsql > \gsql_\rmii{min}$,
  whereas $(\Zgtd)_\rmii{full}$ ranges over the full integration domain (cf.~sec.~\ref{sec:subtr_scheme}).
  The subtraction of subleading $\OO({\rm d}L\,L/L^{2})$ divergences still contains
  a logarithmic dependence on the IR cutoff at the respective $\gsqlmin^{ }$, which will be removed when supplied with four-dimensional results.
  The data for $\Zgtd$ have three errors:
  the Monte-Carlo integration causes an error for the integration range in which lattice points are available (first bracket),
  the error for the integration of the tail caused by the Monte-Carlo error of the points to which the functional form is fitted (second bracket), and
  finally the error introduced by the finite difference integration via the trapezoidal rule (third bracket).%
}
\label{tab:sim_res}
\end{table}

\subsection{Lattice vs.\ perturbation theory}

At $T=100$~GeV and $\Nf=5$, we expect to be deeply in the perturbative regime. Therefore, this serves as a good starting point for comparing our perturbative results to the updated lattice data in tab.~\ref{tab:sim_res}.
\begin{figure}[t]
\centering
  \includegraphics[width=.5\textwidth]{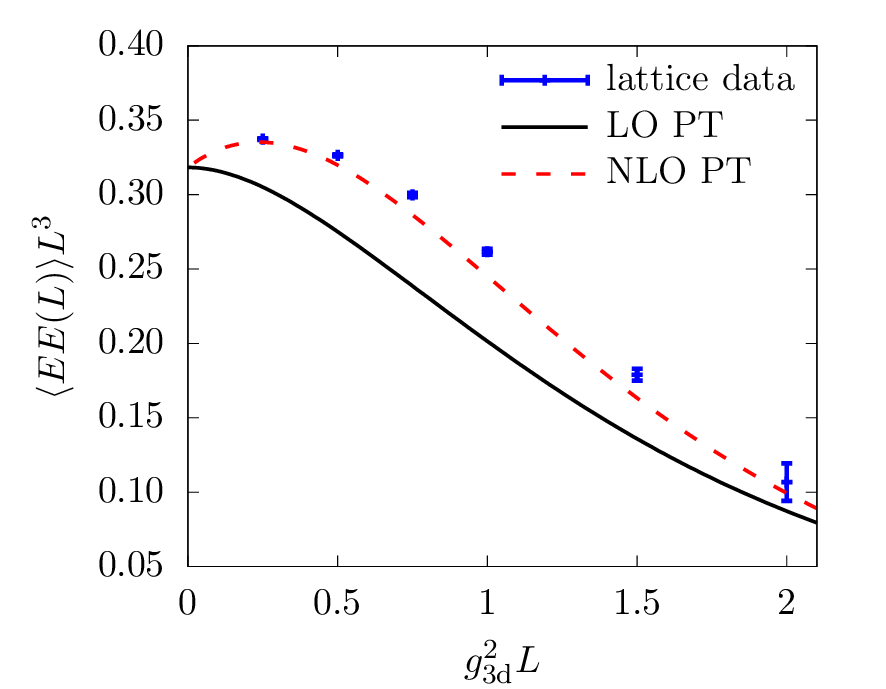}%
  \includegraphics[width=.5\textwidth]{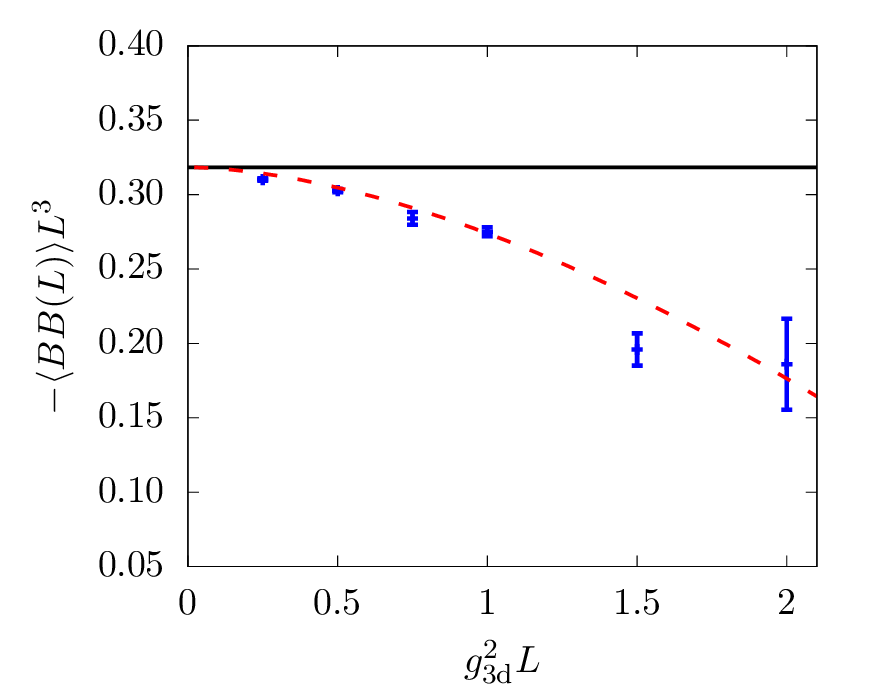}
\caption{%
    Continuum-extrapolated
    $\EE L^3$ (left) and
    $-\BB L^3$ (right)
    from
    leading-order perturbative (LO PT),
    next-to-leading-order perturbative (NLO PT), and
    lattice data
    at $T=100~\mathrm{GeV}$ and $\Nf=5$
    (cf.~tab.~\ref{tab:sim_res}). Correlators are multiplied 
    by $(\gE^2 L)^{3}$ to balance the leading divergence.
    }
\label{fig:comp_LO_NLO_latt}
\end{figure}

Figure~\ref{fig:comp_LO_NLO_latt}~(left) shows that our $\EE$ lattice measurements agree with the NLO analytical result within few multiples of the error up to
separations of $\gsql=2.0$.
Although the LO estimate was already quite close to the analytical data,
the NLO shifted it by a small margin on top of the non-perturbative solution.
The relative correction induced by the NLO is small,
so it seems reasonable to assume good convergence of
the perturbative series at small $\gsql$.

The same holds for $\BB$ in fig.~\ref{fig:comp_LO_NLO_latt}~(right). In fact, the agreement of the lattice data with the perturbative result is surprisingly good, as one would expect contributions from the generically non-perturbative magnetic sector to affect $\BB$ the most of the three correlators.

The correlator $\EB$ vanishes at leading order.
Therefore an analysis of its convergence would require a NNLO result, which is not available as of now.

\begin{figure}[t]
\centering
\includegraphics[width=.5\textwidth]{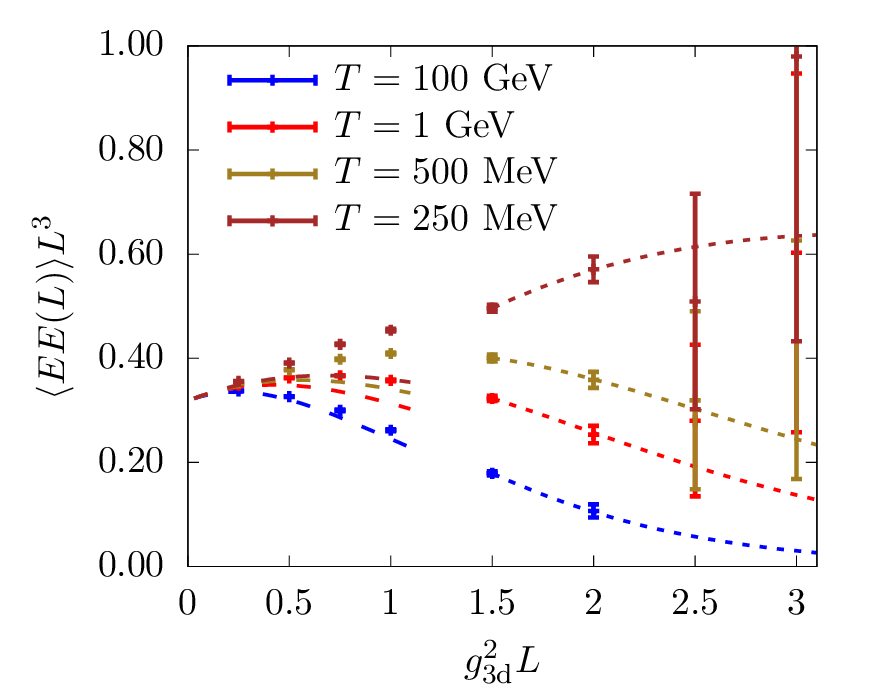}%
\includegraphics[width=.5\textwidth]{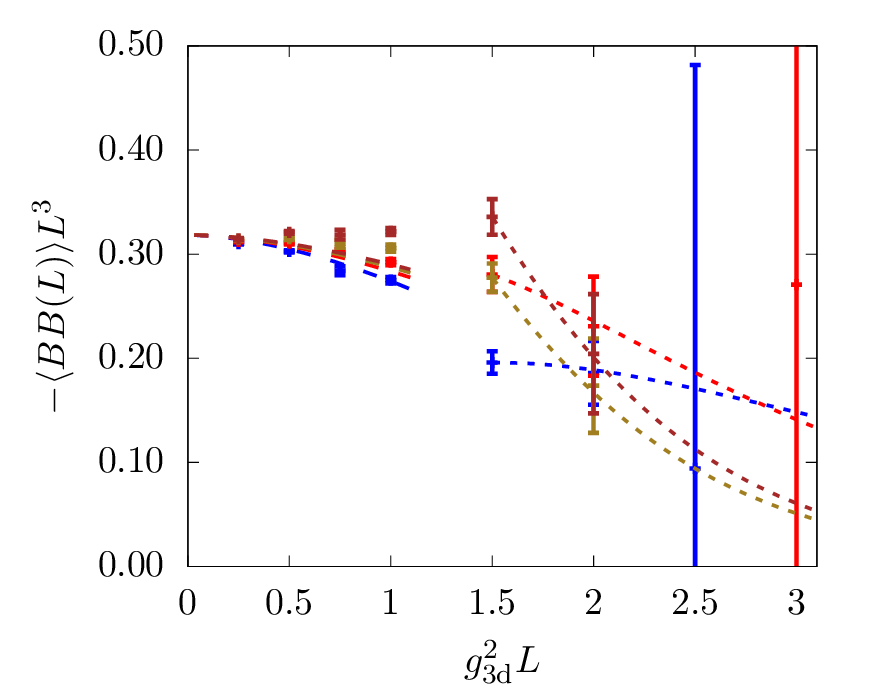}\\
\includegraphics[width=.5\textwidth]{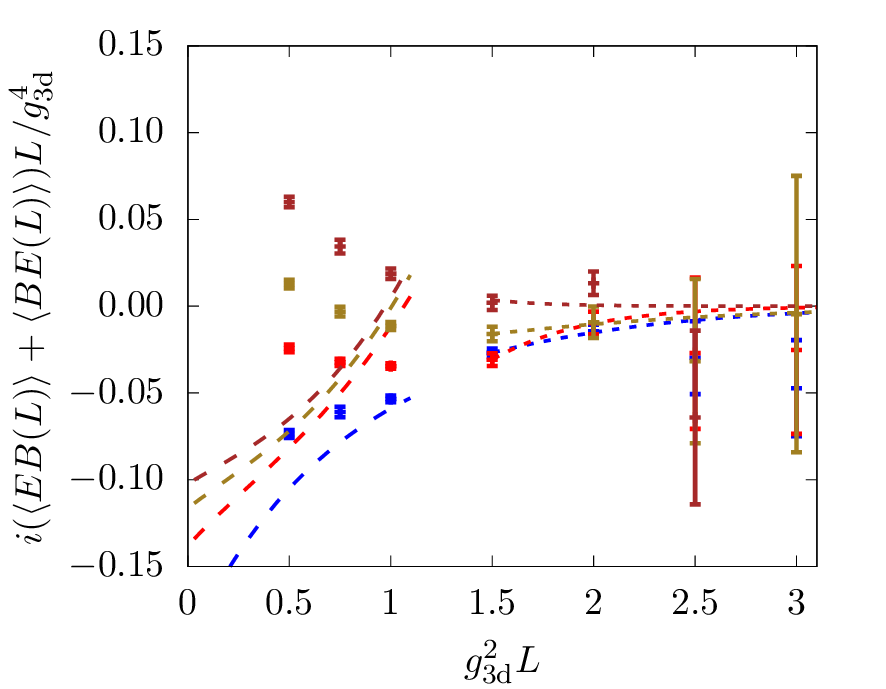}%
\includegraphics[width=.5\textwidth]{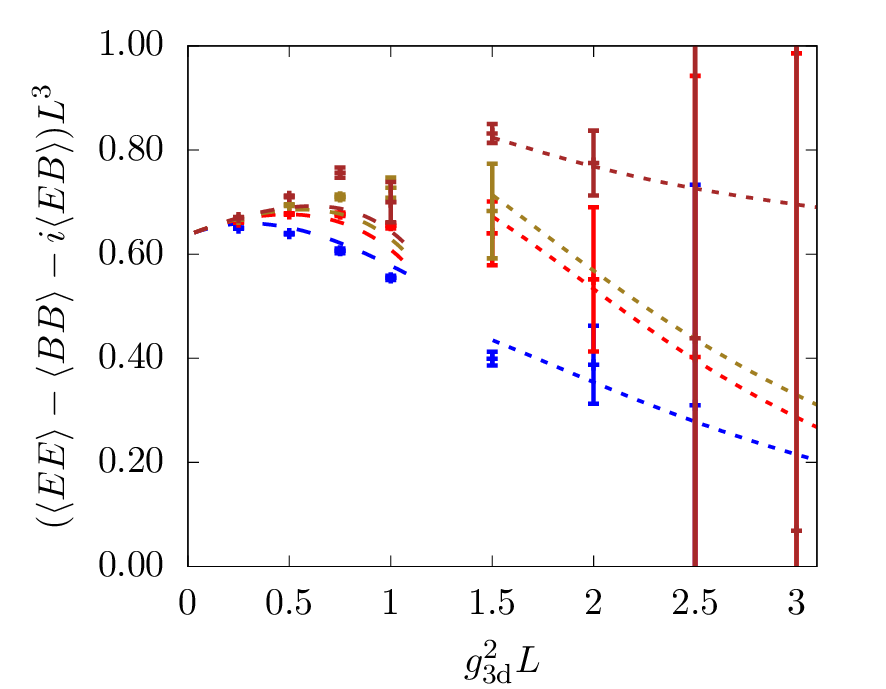}\\
\caption{%
    Continuum-extrapolated
    EQCD lattice data on
    the three separate correlators
    $\EE L^3$,
    $-\BB L^3$,
    $\EB L / \gE^4$,
    (cf.~tab.~\ref{tab:sim_res}) and
    the ${\rm d}L\,L$ integrand of eq.~\eqref{Zg_EQCD}
    with
    modelled long $L$-tail (short dashes) and our 
    NLO perturbative estimate (long dashes).  
    Powers of $\gsql{}$ balance the expected leading 
    divergence of the respective correlator, again.
}
\label{fig:latt_EE_BB_EB}
\end{figure}

Figure~\ref{fig:latt_EE_BB_EB} collects lattice data at all four pairs of $(T,\Nf)$ and
includes their corresponding NLO predictions.
It can be seen that with smaller temperatures -- and larger coupling --
the onset of perturbative behavior in the UV decreases to smaller $\gsql$.
Figure~\ref{fig:latt_EE_BB_EB} also shows that perturbation theory qualitatively predicts well $\EB$ at small $\gsql$ and high temperatures where the coupling is small due to asymptotic freedom. With lower temperatures and larger separations the agreement with our lattice data becomes gradually worse until perturbation theory even fails to predict the correct sign of $\EB$ at small separations and the smallest two temperatures.
What saves the day is that on the one hand,
$\EB$ is numerically suppressed compared to $\EE$ and $\BB$, so its overall impact on $\Zgtd$ is not large,
as seen from fig.~\ref{fig:latt_EE_BB_EB}(lower right).
On the other hand, the tree-level contribution for $\EB$ vanishes.
Therefore, we evaluate only one non-trivial order for $\EB$ and cannot
analyze how well the perturbative series converges or
assess the quality of our perturbative estimate for $\EB$.
Bearing this reasoning in mind,
we accept the mismatch of the lattice data for $\EB$ and
its perturbative prediction in fig.~\ref{fig:latt_EE_BB_EB}.

To mitigate the discretization effects of the numerical integration via the trapezoid rule,
we add and subtract our NLO result to the integration
\begin{equation}
    \int {\rm d}L L \; (\mathrm{corr})_\rmii{latt} =
    \int {\rm d}L L \; \big( (\mathrm{corr})_\rmii{latt} - (\mathrm{corr})_\rmii{NLO} \big)
  + \int {\rm d}L L \; (\mathrm{corr})_\rmii{NLO}
  \, ,
\end{equation}
with the NLO form of the correlators in \eqref{PT_EE_final}, \eqref{PT_BB_final}, and \eqref{PT_EBBE_final}. The last integral can be done analytically, whereas the difference in the first integral is numerically much better behaved:
$\OO\left( \gfour / L \right)$ instead of
$\OO \left( 1 / L^3 \right)$ for $\EE$ and $\BB$, or $\gsq / L^2$ for $\EB$, respectively.
At this point, we would like to caution the reader that this procedure is purely motivated to accelerate numerical convergence and has no physical meaning, in contrast to the subtractions in the following subsection.

As elaborated in~\cite{Moore:2020wvy}, it is necessary to model the large-$\gsql$ tail of the correlators in order to perform the ${\rm d}L\,L$ integration up to $\infty$.
For $\EE$ and $\BB$, the functional form, motivated by~\cite{Laine:1997nq}, is
\begin{equation}
  \frac{A}{\bigl( \gsql \bigr)^2} \exp(-B \cdot \gsql)
  \;,
\end{equation}
with the fitting constants $A$ and $B$.
Considering $\EB$, we find that the data rather follows
\begin{equation}
  A' \exp ( -B' \cdot \gsql )
  \;,
\end{equation}
with the respective fitting constants $A'$ and $B'$.
As already argued above, the impact of $\EB$ on $\Zg$ is small.
Also, the error associated with the large-$\gsql$ tail given in tab.~\ref{tab:sim_res} amounts to roughly $100\%$ of the contribution of the large-$\gsql$ tail.
Therefore, one can exercise a bit of freedom in the choice of the large-$\gsql$ tail modeling.
Moreover, we found that switching between
an effective asymptotic description and
our lattice data was best performed at $\gsql=3.0$ for $\EE$ and $\BB$, but
for the fit of $\EB$ it was most stable when neglecting the noisy $\gsql=3.0$ points and 
switching already at $\gsql=2.5$.

%
\subsection{Subtraction scheme}
\label{sec:subtr_scheme}
As explained in sec.~\ref{sec:matching}, we need to subtract
the UV-divergent limiting behavior from the EQCD calculation.
That is because, in dealing with the contribution of the scale $T$ at
$\mathcal{O}(g^0)$ and
$\mathcal{O}(g^2)$ (which we have not computed yet),
it is impractical to separate the contribution
of the Matsubara zero mode from that of the other modes. Hence, that 
$T$-scale contribution ends up containing a zero-mode contribution
computed without any resummation of the screening mass: 
it thus must agree with the $\mD\to 0$ limit of the EQCD calculation. Hence,
this  $\mD\to 0$  or UV limit of the resummed EQCD calculation must
be subtracted from our non-perturbative evaluation, so that the latter 
could, in our upcoming paper, be summed with the contribution 
of the scale $T$ at order $\mathcal{O}(g^2)$.

This subtraction poses no problem for the leading $1/L^3$ UV behavior
in eq.~\eqref{eq:LO:UV},
which can be safely integrated in ${\rm d} L\, L$ up to infinity.
As we argued,
that is not the case for the subleading $1/L^2$ terms
in eq.~\eqref{eq:NLO:UV},
which are IR divergent. 
We introduce an IR cutoff for the NLO subtraction, to be absorbed
once the $T$-scale contribution becomes available.
For convenience, we set this cutoff to
the respective smallest lattice separation $\gsqlmin^{ }$ of each correlator, i.e.\
$0.25$ for $\EE$ and $\BB$, and
$0.5$ for $\EB$; see ~tab.~\ref{tab:sim_res}.
Therefore, we modify the values for the correlators in tab.~\ref{tab:sim_res} by subtracting
the leading UV counterterm~\eqref{eq:LO:UV} over the entire range, and
add the NLO perturbative result, subtracted by the NLO divergence, at $\gsql < \gsqlmin^{ }$, {\em viz.}
\begin{equation}
\label{def_zgct}
    (\Zgtd)_\rmii{full} \equiv
    (\Zgtd)_\rmii{cutoff}
  + (\Zgtd)_\rmii{UV}
  \;.
\end{equation}
The precise form of the
UV completions (in EQCD, i.e.\ unphysical)
$(\Zgtd)_\rmii{UV}$
is given in
tab.~\ref{modification_terms}.
Bear in mind that the LO subtractions
$\EE_\rmii{LO}^\rmii{subtr}$ and
$\BB_\rmii{LO}^\rmii{subtr}$ are already factored into
$\left( \Zgtd \right)_\rmii{cutoff}$ up to the UV cutoff $\gsqlmin^{ }$.
\begin{table}[t]
\centering
\begin{tabular}{|c||c|c|}
  \hline
  &
  $\bigl[0,\gsqlmin^{ }\bigr)$ &
  $\bigl[\gsqlmin^{ },\infty\bigr)$ \\
  \hline
  $\EE$ &
  $\EE_\rmii{LO} - \EE_\rmii{LO}^\rmii{subtr}
  +\EE_\rmii{NLO} - \EE_\rmii{NLO}^\rmii{subtr}$ &
  $
  -\EE_\rmii{LO}^\rmii{subtr}$ \\
  $\BB$ &
  $\BB_\rmii{NLO}$ &
  $
  -\BB_\rmii{LO}^\rmii{subtr}$ \\
  $\EB$ &
  $\EB_\rmii{NLO}$ & --\\
  \hline
\end{tabular}
\caption{%
  Definition of  $(\Zgtd)_\rmii{UV}$, as given in eq.~\eqref{def_zgct}. 
  The second and third columns give the integrands for the three contributions --
  which are to be added as per eq.~\eqref{Zg_EQCD} --
  in different intervals of $\gsql$ and with IR cutoff $\gsqlmin^{ }$.
  We recall that at leading order
  $\BB_\rmii{LO}$ agrees with its subtraction and
  $\EB_\rmii{LO}$ vanishes.
  Moreover, $\BB_\rmii{NLO}^\rmii{subtr}$ and $\EB_\rmii{NLO}^\rmii{subtr}$ 
  vanish since the respective correlators are UV-finite.
  }
\label{modification_terms}
\end{table}
The integrated UV completions can be found in
tab.~\ref{modification_values}.
The final results of this paper, the gauge condensate with all necessary subtractions of unresummed zero-mode-contributions, and the IR divergence induced by the $1/L^2$-subtraction regulated by the respective $\gsqlmin^{ }$, are displayed as $(\Zgtd)_\rmii{full}$ in tab.~\ref{tab:sim_res}.
\begin{table}[ht]
\centering
\begin{tabular}{|c||c|c|c|c|}
  \hline
  $(T,\Nf)$ &
  $(250~\mathrm{MeV},3)$ &
  $(500~\mathrm{MeV},3)$ &
  $(1~\mathrm{GeV},4)$ &
  $(100~\mathrm{GeV},5)$ \\
  \hline
  $\int\! {\rm d} L\,L\,\EE / \gsq$ &
  $-0.0316$ &
  $-0.0386$ &
  $-0.0505$ &
  $-0.0881$
  \\
  $\int\! {\rm d} L\,L\,\BB / \gsq$ &
  $\hp{-}0.0110$ &
  $\hp{-}0.0123$ &
  $\hp{-}0.0143$ &
  $\hp{-}0.0192$
  \\
  $\int\! {\rm d} L\,L\,\EB / \gsq$ &
  $-0.0435$ &
  $-0.0485$ &
  $-0.0556$ &
  $-0.0724$
  \\
  \hline
\end{tabular}
\caption{%
    UV completions according to tab.~\ref{modification_terms}
    integrated over range $\big[ 0, \, \gsqlmin^{ } \big)$.
  }
\label{modification_values}
\end{table}

Revisiting tab.~\ref{tab:sim_res} after these subtractions, we discover that all values of $\left( \Zgtd \right)_\rmii{full}$ except the $T=250$~MeV one are consistent with zero within errors. This stresses once more the good quality of our perturbative result and the small deviations of our lattice result from the perturbative curves in the part of the integration range that dominates $\Zgtd$. The final reasoning and intuition about $\Zgtd$ will be postponed to a forthcoming publication where the four-dimensional perturbative calculation is carried out~\cite{wip}.

%
\section{Discussion}
\label{sec:discussion}

The correlator of two insertions of the color-force is a crucial ingredient for the computation of asymptotic jet masses. In particular, it receives non-perturbative contributions from the magnetic and electrostatic sector of hot thermal QCD. In the present paper, we provided a twofold calculation of this correlator in EQCD:
an analytical calculation to next-to-leading order as well as
a non-perturbative simulation of all involved correlators.

The NLO calculation is needed for the final goal of integrating the force-force correlator from zero to infinite separation.
On the one hand, it matches well at small separations with the non-perturbative lattice data for the numerically dominant contributions.
Therefore, the NLO result improves the modelling of the small $\gsql$ region,
which is inaccessible to lattice simulations.
On the other hand, it allows to understand the nature of UV behavior of the EQCD result.
On dimensional grounds,
the NLO in EQCD is the last order which can give rise to UV divergent terms in $\Zgtd$.
These divergences need to be subtracted to avoid double-counting of
degrees of freedom in EQCD and in the outstanding parts of full QCD.
Ultimately, our goal is to subtract the UV limit from the analytical EQCD calculation,
since we know that EQCD is unphysical in the UV, and replace it with the correct
UV behavior obtained from full (perturbative) QCD.

We found that presenting the complete EQCD result seemed a natural part of the required work, which has to be finally supplied with the aforementioned  perturbative calculation of the contribution of the scale $T$ at $\mathcal{O}(g^2)$ in full QCD to deliver a meaningful non-perturbative result for $\Zg$ and consequently $\mIsq$.
In particular, having the full QCD result at hand allows one to
remove the unphysical UV behavior of EQCD, which has been recast by
the subtraction  procedure of sec.~\ref{sec:subtr_scheme}
as a dependence on an intermediate regulator $\gsqlmin^{ }$.

In an upcoming paper, we will deal with that missing contribution by determining $\Zg$ at NLO in a naive perturbative expansion in full QCD, which
is appropriate for capturing the contribution of the scale $T$. We anticipate
that practicality concerns will most likely force us to tackle such a
computation in momentum space, forgoing the possibility of keeping
the $L$ dependence in intermediate steps. Hence, the expected IR divergence
would show up at small momenta, and we would need to translate that in
the $L$-space form of sec.~\ref{sec:subtr_scheme}. That would lift
the logarithmic, unphysical dependence on $\gsqlmin^{ }$ introduced by
eq.~\eqref{def_zgct}, leading to a finite, scheme-independent
result for the asymptotic mass which incorporates the IR EQCD contribution to
all orders.

%
\section*{Acknowledgements}
We would like to thank
Eamonn Weitz,
Alexander M.~Eller,
Aleksi~Vuorinen, and
Kari~Rummukainen for useful conversations and fruitful suggestions.
We thank Zavolan Lab at the Biozentrum of the University of Basel, where PS was hosted during the later stages of this work.
JG acknowledges support by a PULSAR grant from the R\'egion Pays de la Loire.
GM was supported by the Deutsche Forschungsgemeinschaft (DFG, German Research Foundation) --
project number 315477589 --
CRC TRR 211.
PS was supported
by the European Research Council, grant no.~725369, and
by the Academy of Finland, grant no.~1322507.
NS acknowledges support from Academy of Finland grants 267286 and 320123.

%
\appendix

%
\section{NLO contributions}
\label{sec:NLO:details}

This appendix collects the integrals and technical details of
the contributions of fig.~\ref{fig:diagrams}
leading up to the next-to-leading order contribution of
the $\Zgtd$ correlators in sec.~\ref{sec:NLO}.
In the arising diagrams,
a curly line denotes a spatial gauge boson ($A_i$) and
a solid line an adjoint scalar ($\Phi$).
If not explicitly stated otherwise, all results are collected in Feynman gauge
with $(\xi=1)$ in the gauge propagator~\eqref{eq:gl:prop:mom}.
For convenience, we perform calculations in
the adjoint representation which corresponds to twice
the results in the fundamental representation~\eqref{eq:adj:fund}.

%
\subsection{Master integrals}
\label{sec:masters}

With momenta
$p=(\vec{p}_{\perp}^{ },p_{z}^{ })$,
we regularize integrals in $d=3-2\epsilon$ dimensions employing 
the $\overline{\mbox{MS}}$ renormalization scale
$\bmu^{2} = 4\pi e^{-\gammaE}\mu^{2}$ and
$
\int_{\vec{p}}\equiv
\mu^{3-d}\int\frac{{\rm d}^d \vec{p}}{(2\pi)^d}
$,
$
\int_{\vec{p}_{\perp}}\equiv
\mu^{3-d}\int\frac{{\rm d}^{d-1} \vec{p}_{\perp}^{ }}{(2\pi)^{d-1}}
$.
Thus, the position space
free massless and massive propagators with general powers (cf.~sec.~\ref{sec:LO})
follow from the Fourier transform:
\begin{align}
\label{FT:psq}
\mathcal{I}_{\alpha;0}(r) &=
  \int_{\vec{p}}
  \frac{\e^{i\vec{p}\cdot\vec{r}}}{\big[p^2\big]^{\alpha}} =
  \Big( \frac{\bmu^{2}e^\gammaE}{4\pi} \Big)^\epsilon
  2^{d-2\alpha}
  \frac{[r^2]^{\alpha - \frac{d}{2}}}{(4\pi)^\frac{d}{2}}
  \frac{
    \Gamma(\frac{d}{2} - \alpha)}{
    \Gamma(\alpha)}
  \;,\\
\label{FT:psq:m}
\mathcal{I}_{\alpha;m}(r) &=
  \int_{\vec{p}}
  \frac{\e^{i\vec{p}\cdot\vec{r}}}{\bigl[p^{2} + m^{2}\bigr]^{\alpha}} =
  \Big( \frac{\bmu^{2}e^\gammaE}{4\pi} \Big)^\epsilon
  \frac{2^{\frac{d}{2}+1-\alpha}}{(4\pi)^\frac{d}{2}}
  \Bigl[\frac{m}{r}\Bigr]^{\frac{d}{2} - \alpha}
  \frac{
    K_{\frac{d}{2} - \alpha}(m r)}{
    \Gamma(\alpha)}
\;.
\end{align}
Here
$K_{\nu}(z)$ is the modified Bessel functions of second kind and
coordinates are chosen so that $\vec{r}$ points in the $z$-direction
with $r=|\vec{r}|$.

%
\subsection{Diagram $(b)$}
\label{sec:b}

The simplest NLO contribution is diagram $(b)$
which is depicted and calculated as follows:
\begin{align}
\label{eq:b:EE}
  \ToptWBB(\Lsri,\Asai,\Asai) &+
  \ToptWBB(\Lsri,\Asai,\Agliii) =
  2 \times (b)^{\rmii{EE}}
  \nn &=
  2\gsq\CA^{2}\CF^{ }
  \int_{\vec{pk}}
  \frac{e^{ip_{z}L}}{[p^2 + \mD^2]}
  \frac{(p^2 - p_{z}^2)}{2}
  \int_0^L \! {\rm d} z_1
  \int_{z_1}^L \! {\rm d} z_2\,
  e^{ik_{z}(z_{1} - z_{2})}\Bigl[
      \frac{1}{k^2 + \mD^2}
    - \frac{1}{k^2}
  \Bigr]
  \nn[2mm] &=
  - \frac{2\gsq\CA^2 \CF}{(4\pi L)^{2}} \frac{\e^{-\mD L}(1 + \mD L)}{\mD L}
  \nn &\hp{{}=} \times
  \Big[
    1
    - \e^{-\mD L}
    - \mD L\bigl(
      1 - \gammaE - \ln(\mD L) - E_{1}(\mD L)
  \bigr)
  \Bigr]
  \;, \\[3mm]
  \ToptWBB(\Lsri,\Agliii,\Asai) &+
  \ToptWBB(\Lsri,\Agliii,\Agliii) =
  2 \times (b)^{\rmii{BB}}
  \nn &=
  \frac{2 \gsq \CA^2 \CF}{(4\pi L)^2}\frac{1}{\mD L}
  \nn &\hp{{}=} \times
  \Big[
    1
    -\e^{-\mD L}
    - \mD L\bigl(
      1 - \gammaE - \ln(\mD L) - E_{1}(\mD L)
  \bigr)
  \Big]
  \;,
\end{align}
using
the free propagator integrals~\eqref{FT:psq} and \eqref{FT:psq:m}
and the exponential integral function with respective limits:
\begin{equation}
\label{eq:E1}
  E_{1}(x) \equiv \int_{x}^{\infty} \!{\rm d}t\, \frac{\e^{-t}}{t}
  \;,\quad
  E_{1}(x) \stackrel{x\gg 1}{\approx} \frac{\e^{-x}}{x}
  \;,\quad
  E_{1}(x) \stackrel{x\ll 1}{\approx} \ln\frac{1}{x} - \gammaE
  \;,
\end{equation}
with
$E_{1}(x) = -\mbox{Ei}(-x)$ for $x > 0$.
In both contributions the corresponding LO operator interaction of
diagram \hyperref[sec:LO]{$(a)$} factorizes from
the internal propagator between the two Wilson line insertions.
As exemplified in the internal $z_{1,2}$ integration of
$(b)^{\rmii{EE}}$ in eq.~\eqref{eq:b:EE},
a UV divergence of individual integrals is of opposite sign which renders
the overall $\EE$ and $\BB$ contributions separately finite.

%
\subsection{Diagram $(c)$}\label{sec:c}

As depicted in fig.~\ref{fig:diagrams}, diagram $(c)$ requires
the self-energies of
the adjoint scalar $\Phi^{a}$ and
the spatial gauge bosons $A^{a}_{i}$;
diagrammatically given by
\begin{align}
  \TopSi(\Lsa3,1) &=
   \TopoST(\Lsa3,\Agl3)
  +\TopoST(\Lsa3,\Asa3)
  +\TopoSB(\Lsa3,\Agl3,\Asa3)
  \;,\\
  \TopSi(\Lge3,1) &=
   \TopoST(\Lge3,\Agl3)
  +\TopoST(\Lge3,\Asa3)
  +\TopoSB(\Lge3,\Agl3,\Agl3)
  +\TopoSB(\Lge3,\Asa3,\Asa3)
  -\TopoSB(\Lge3,\Ahg3,\Ahg3)
  \;,
\end{align}
where
a curly line is a spatial gauge boson,
a solid line an adjoint scalar, and
a dotted directed line a ghost field.
The analytical forms of the self-energies are
\begin{eqnarray}
\label{eq:SE:phi:mom}
\Pi_{\Phi^{a}\Phi^{b}}(p^2) &=&
  \delta^{ab}\gE^{2}\CA^{ }\biggl\{
    \int_{\vec{q}} \frac{1}{[q^2 + \mD^{2}]}\Bigl[
      (d-1) - (d-2)\xi
      + \frac{6\CF - \CA}{24\CA}\frac{\lambdaE}{\gE^{2}}
      \Bigr]
  \nn[2mm] &&
    -\int_{\vec{q}}\frac{p^2 - \mD^{2}}{[q^2 + \mD^{2}](p-q)^2}\Bigl[
    (d-1) - (d-3)\xi
  \Bigr]
\biggr\}
  \\[2mm] &\stackrel{d=3-2\epsilon}{=}&
\label{eq:SE:phi}
  \delta^{ab}\frac{\gE^{2}\mD^{ }\CA^{ }}{16\pi}\Bigl(
      4(\xi - 2)
    - \frac{6\CF - \CA}{6\CA}\frac{\lambdaE}{\gE^{2}}
    - 8\frac{p^2 - \mD^{2}}{p\,\mD}\arctan\Bigl(\frac{p}{\mD}\Bigr)
    \Bigr)
    \;,\hspace{1cm}\\[3mm]
\label{eq:SE:A:mom}
\Pi^{\rmii{T}}_{A^a_i A^b_j}(p^2) &=&
  \delta^{ab}\frac{\gE^{2}\CA^{ }}{(d-1)}\biggl\{
    \int_{\vec{q}} \frac{(d-2)}{[q^2 + \mD^{2}]}
  + \frac{1}{2}\int_{\vec{q}}\frac{p^2 + 4\mD^{2}}{[q^2 + \mD^{2}][(p-q)^2+\mD^2]}
  \nn[2mm] &&
  - \int_{\vec{q}}\frac{p^2}{q^2 (p-q)^2}\Bigl[
    \frac{7d^2 - 19d + 16}{8}
    + (d-1)\frac{4(\xi + 5) - d(\xi + 6)}{8}\xi
    \Bigr]
  \biggr\}
  \\[2mm] &\stackrel{d=3-2\epsilon}{=}&
\label{eq:SE:A}
\delta^{ab}\frac{\gE^{2}\CA^{ }}{16\pi}\Bigl(
    - 2\mD
    - \frac{11 + (\xi + 2)\xi}{4}\pi p
    + \frac{p^2 + 4\mD^2}{p}\arctan\Bigl(\frac{p}{2\mD}\Bigr)
    \Bigr)
\;,
\end{eqnarray}
where
the gauge boson self-energy, 
$\Pi^{ }_{A^a_i A^b_j}(p^2) =
\bigl(\delta_{ij} - \frac{p_i p_j}{p^2}\bigr)
\Pi^{\rmii{T}}_{A^a_i A^b_j}(p^2)$,
is transverse in three dimensions.
Addtionally, we
set $d=3-2\epsilon$ and
employ the master integral~\cite{Giovannangeli:2005rz}
\begin{align}
\label{eq:arctan}
  \int_{\vec{q}}
  \frac{1}{
    [q^2 + m_{1}^2]
    [(p-q)^2 + m_{2}^2]}
  = \frac{1}{4\pi p}\arctan\Bigl(\frac{p}{m_{1}+m_{2}}\Bigr)
  \;,\quad
  {\rm for}\;
  m_{1}+m_{2} > 0
  \;.
\end{align}
The gauge boson self-energy agrees with
\cite{Laine:2005ai} in the limit of soft external momenta $\frac{p^2}{\mD^2}\ll 1$ and
\cite{CaronHuot:2008ni} in Feynman gauge ($\xi= 1$).
For the latter one needs to account for an additional relative minus sign between
spatial and temporal correlators.

Inserting the self-energies and sources of the $\EE$ and $\BB$ correlators
{\`a} la eq.~\eqref{eq:LO:int} in their corresponding diagrams yields
in momentum space
\begin{eqnarray}
\ToptWSBB(\Lsri,\Asai,\Asai,\Asai,\Asai)
&=&
2 \times (c)^{\rmii{EE}} =
  \frac{1}{2}
  \int_{\vec{p}} \frac{e^{ix_{z}p_{z}}}{[p^2 + \mD^2]^{2}}\bigl(p^2 - p_{z}^2\bigr)
  \,\delta^{ab}\,\bigl[-\Pi^{ }_{\Phi^{a}\Phi^{b}}(p^2)\bigr]
  \;, \\
\ToptWSBB(\Lsri,\Agliii,\Agliii,\Asai,\Asai)
&=&
2 \times (c)^{\rmii{BB}} =
  \frac{1}{2}
  \int_{\vec{p}} \frac{e^{ix_{z}p_{z}}}{[p^2]^{2}}\bigl(p^2 + (d-2) p_{z}^2\bigr)
  \,\delta^{ab}\,\bigl[-\Pi^{\rmii{T}}_{A^a_i A^b_j}(p^2)\bigr]
\;.
\end{eqnarray}
The emergent integrals are again
free propagators according to appendix~\ref{sec:masters}.
To reduce the number of integrals,
one can make use of integration-by-parts identities
$\int_{\vec{p}_\perp}\frac{{\rm d}}{{\rm d} \vec{p}_\perp} \vec{p}_\perp
\frac{p_{z}^\beta}{[p_{z}^2 + p_{\perp}^2 + m^2]^{\alpha}} = 0$
among transverse momenta in 
$p^2 = p_{\perp}^{2} + p_{z}^{2}$
\begin{equation}
\label{eq:ibp:1l}
  \int_{\vec{p}_{\perp}}
  \frac{p_{z}^{\beta+2}}{[p^2 + m^2]^{\alpha+1}} =
    \frac{(2\alpha-d+1)}{2\alpha}
    \int_{\vec{p}_\perp} \frac{p_{z}^{\beta}}{[p^2 + m^2]^{\alpha}}
    - 2m^2\int_{\vec{p}_\perp} \frac{p_{z}^{\beta}}{[p^2 + m^2]^{\alpha+1}}
  \;.
\end{equation}
The remaining integrals needed for the $p$-integration of
eqs.~\eqref{eq:SE:phi} and \eqref{eq:SE:A}
follow by inserting eq.~\eqref{eq:arctan}
in the following
\begin{align}
\mathcal{I}_{\alpha;mm0}^{ } &=
  \int_{\vec{p}} \frac{e^{i\vec{p}\cdot\vec{r}}}{
    [p^2 + m^2]^{\alpha}
  }
  \int_{\vec{q}}
  \frac{1}{
    [q^2 + m^2]
    (p-q)^2}
  \;,\\
\label{eq:I:a:0mm}
\mathcal{I}_{\alpha;0mm}^{ } &=
  \int_{\vec{p}} \frac{e^{i\vec{p}\cdot\vec{r}}}{
    [p^2]^{\alpha}
  }
  \int_{\vec{q}}
  \frac{1}{
    [q^2 + m^2]
    [(p-q)^2 + m^2]}
  \;,\\
\mathcal{I}_{0;mm0}^{ } &=
  \frac{e^{-mr}}{(4\pi r)^2}
  \;,\\
\mathcal{I}_{1;mm0}^{ } &=
  \frac{e^{-mr}}{(4\pi)^2}\frac{1}{2mr}\Bigl[
    \ln(2mr) + \gammaE + e^{2mr}E_{1}(2mr)
    \Bigr]
  \;,\\
\mathcal{I}_{2;mm0}^{ } &=
  \frac{e^{-mr}}{(4\pi m)^2}\frac{1}{4mr}\Bigl[
      (1+mr)\bigl(\ln(2mr) + \gammaE\bigr)
    + (1-mr)e^{2mr}E_{1}(2mr)
  \Bigr]
  \;,\\
\mathcal{I}_{0;0mm}^{ } &=
  \frac{e^{-2mr}}{(4\pi r)^2}
  \;,\\
\mathcal{I}_{1;0mm}^{ } &=
  \frac{1}{(4\pi)^2}\frac{1}{2mr}\Bigl[
    1
    - e^{-2mr}
    + 2mr E_{1}(2mr)
  \Bigr]
  \;,\\
\partial^2_{r}\,\mathcal{I}_{2;0mm}^{ } &=
  \frac{1}{(4\pi)^2}\frac{1}{12(mr)^3}\Bigl[
    -1
    + e^{-2mr}\bigl(1 + 2mr + 2(mr)^2\bigr)
    - 4(mr)^3 E_{1}(2mr)
  \Bigr]
  \;,
\end{align}
which we evaluated using the residue theorem along
the integration contour as in~\cite{Nadkarni:1986cz}.
For the last integral
$\mathcal{I}_{2;0mm}^{ }$
only its finite second $r$-derivative is needed.
As a result,
we obtain for the diagonal contributions in Feynman gauge:
\begin{eqnarray}
\label{eq:c:EE}
2 \times (c)^{\rmii{EE}}
&=&
  \frac{2\gE^{2}\CF^{ }\CA^{2}}{(4\pi L)^{2}} e^{-\mD L}\Big[
      2
    +\mD L \frac{6\CF - \CA}{48\CA}\frac{\lambdaE^{ }}{\gE^{2}}
  \nn[3mm] &&\hp{\frac{2\gE^{2}\CF^{ }\CA^{2}}{(4\pi L)^{2}} e^{-\mD L}\Big[+}
    + \mD L\Big( \frac{3}{2} - \gammaE - \ln(2\mD L) - e^{2\mD L} E_{1}(2\mD L)\Big)
  \Big]
  \\[2mm] &\stackrel{L\to 0}{=}&
    \frac{2\gE^{2}\CF^{ }\CA^{2}}{(4\pi L)^{2}}
    \biggl[
      2
    - \mD L \frac{24 \CA - x (6\CF - \CA) }{48\CA}
  \nn[2mm] &&
  \label{eq:c:EE:UV}
    + (\mD L)^2 \frac{\CA (-240 + 192 \gammaE + 192 \ln (2 \mD L)) -2 x ( 6 \CF - \CA)}{96\CA} \biggr]
    + \OO(L) \;,\hspace{1cm}
    \\[3mm]
\label{eq:c:BB}
2 \times (c)^{\rmii{BB}}
&=&
  -\frac{2\gsq\CA^{2}\CF^{ }}{(4\pi L)^2}\frac{1}{12\mD L}\Bigl[
    e^{-2\mD L}
    \bigl(1+2\mD L - 4(\mD L)^2\bigr)
  \nn &&\hp{-\frac{2\gsq\CA^{2}\CF^{ }}{(4\pi L)^2}\frac{1}{12\mD L}\Bigl[}
    - 1 + 8(\mD L)^3\,E_{1}(2\mD L)
  \Bigr]
  \\ &\stackrel{L\to 0}{=}&
  \frac{2\gsq\CA^{2}\CF^{ }}{(4\pi L)^{2}} \Bigl[
      \frac{\mD L}{2}
    + \frac{2(\mD L)^2}{9}\bigl( \ln(2\mD L) + 3\gammaE - 4\bigr)
  \Bigr]
  + \rmO(L)
  \;,
\end{eqnarray}
with
$x=\lambdaE/\gsq$.
To understand their UV asymptotics we expand for small $\mD L\to 0$ which vanishes for
$(c)^{\rmii{BB}}$ in~\eqref{eq:c:BB}.

%
\subsection{Diagrams $(d), (i), (j)$}

Diagrams $(d_1)$, $(d_2)$, $(i),$ and $(j)$ each feature at least one
momentum integration with an odd integrand, which therefore vanishes.
Thus none of these diagrams contribute.

%
\subsection{Diagram $(e)$}

Conveniently diagram $(e)$ is finite.
The scalar contribution originates from
the non-Abelian part of $D_x\Phi$ and
the gauge contribution originates from
the non-Abelian part of the field-strength-tensor
$F^a_{xz} \supset \gE f^{abc} A_x^b A_z^c$.
Their momentum space integral representations are
\begin{align}
  \ToptWS(\Lsri,\Asai,\Agliii) =
  2 \times (e)^{\rmii{EE}} &=
  2\gsq\CA^{2}\CF^{ }
  \frac{(d-1)}{2}
  \int_{\vec{p}}\frac{e^{ip_{z}L}}{[p^2 + \mD^2]}
  \int_{\vec{k}}\frac{e^{ik_{z}L}}{k^2}
  \nn &=
  \frac{2\gsq\CF\CA^2}{(4\pi L)^2} \e^{- \mD L}
  \;,\\[1mm]
  \ToptWS(\Lsri,\Agliii,\Agliii) =
  2 \times (e)^{\rmii{BB}} &=
  2\gsq\CA^{2}\CF^{ }
  \frac{(d-1)}{2}
  \biggl[\int_{\vec{p}}\frac{e^{ip_{z}L}}{p^2}\biggr]^2
  \nn &=
  \frac{2\gsq\CF\CA^2}{(4\pi L)^2}
  \;,
\end{align}
where we implicitly used that
the gluon propagator is spatially diagonal in Feynman gauge.
The result is logarithmically divergent which, however, vanishes
after considering opposing signs of $\EE$ and $\BB$ in~\eqref{Zg_EQCD}.

%
\subsection{Diagrams $(f),(h)$}

Diagram $(f)$ corresponds to the Fourier transform of the EQCD three-point vertices.
Starting from the electric scalar contribution in momentum space integral representation,
the strategy is to first integrate over the $z$-coordinate of
the single $A_z$ field sourced from the Wilson line
\begin{align}
\label{f:EE}
\ToptWM(\Lsri,\Asai,\Asai,\Lgliii) &=
2 \times (f)^{\rmii{EE}}
  \nn &=
  -\gsq\CA^{2}\CF^{ }
  \int_{\vec{pk}} \big[e^{ip_{z}L}-e^{ik_{z}L}\big]\,
    \frac{p_{z} + k_{z}}{p_{z} - k_{z}}
    \frac{
       \vec{p}\cdot\vec{k}
    - p_{z}^{ }k_{z}
      }{
    [p^2 + \mD^2]
    [k^2 + \mD^2]
    (p-k)^2}\nn 
    &=
  -2\gsq\CA^{2}\CF^{ }
  \int_{\vec{pk}} e^{ip_{z}L}\,
   \mathbb{P}\, \frac{p_{z} + k_{z}}{p_{z} - k_{z}}
    \frac{
       \vec{p}_\perp\cdot\vec{k}_\perp
      }{
    [p^2 + \mD^2]
    [k^2 + \mD^2]
    (p-k)^2}
  \;.
\end{align}
In obtaining the final line, we have exploited the
$\vec{p}\leftrightarrow \vec{k}$ symmetry
of the integrand to reshuffle the $e^{ik_zL}$ term into an $e^{ip_zL}$. As
the integrand was, prior to this step, well behaved for $p_z=k_z$, we can treat
the apparent pole at
$p_{z} - k_{z}$ introduced by this reshuffling with a principal value prescription  $\mathbb{P}$ (cf.~\cite{Burnier:2009bk}).
To perform the momentum integrations in eq.~\eqref{f:EE}, we proceed as follows.
We 
\begin{itemize}
  \item[(i)]
    combine the two $\vec{k}$-dependent denominators with a Feynman parameter $x$,
  \item[(ii)]
    perform
    the ${\rm d}^2\vec{k}_\perp$ and
    the $\mathbb{P}$-regulated ${\rm d}k_z$ integrals, 
    followed by the one over the Feynman parameter $x$
    which yields 
\end{itemize}
\begin{align}
\label{f:EE_intermediate}
  2 \times (f)^{\rmii{EE}} =
  - 2\gsq\CA^{2}\CF^{ }
  \int_{\vec{p}} \frac{e^{ip_{z}L}}{p^2 + \mD^2}
  \bigg[&-\frac{p_z \left(\mD^2+p_z^2\right) \tan^{-1}\bigl(\frac{p_z}{\mD}\bigr)}{2\pi 
  \left(\mD^2+p^2\right)}
    - \frac{\mD p_\perp^2}{8\pi p^2}
  \nn &
  + \frac{\left(\mD^2 p_\perp^2-p_\perp^4+3 p_\perp^2 p_z^2+4 p_z^4\right)
  \tan ^{-1}\bigl(\frac{p}{\mD}\bigr)}{8\pi p^3}\bigg]
  \;,
\end{align}
\begin{itemize}
  \item[(iii)]
    perform the ${\rm d}^3\vec{p}$ integrations as in appendix~\ref{sec:c}.    
\end{itemize}
We obtain
\begin{eqnarray}
\label{f:EE2}
2 \times (f)^{\rmii{EE}}
&=&
  - \frac{2\gsq\CA^{2}\CF}{(4\pi L)^2}\frac{e^{-\mD L}}{\mD L}\Big[
        \mD L (4 + \mD L)
  \nn &&\hp{=\frac{2\gsq\CA^{2}\CF}{(4\pi L)^2}\frac{e^{-\mD L}}{\mD L}\Big[}
    - \Bigl(3 + 3\mD L + (\mD L)^2\Bigr)\Bigl(\ln(2\mD L)+\gammaE\Bigr)
  \nn &&\hp{=\frac{2\gsq\CA^2\CF}{(4\pi L)^2}\frac{e^{-\mD L}}{\mD L}\Big[}
    - e^{2\mD L}\Bigl(3 - 3\mD L + (\mD L)^2\Bigr)E_{1}(2\mD L)
  \Big]
  \;, \nn &\stackrel{L\to 0}{=}&
  \frac{2\gsq\CA^{2}\CF}{(4\pi L)^2} \frac{1}{3}\Bigl[6 - 2(\mD L)^2 \Bigr]
  + \OO(L) \;.
\end{eqnarray}
Diagram $(h)^{\rmii{EE}}$ and its mirrored partner vanish
because each transverse-momentum integration has an odd integrand,
\begin{align}
  \ToptWal(\Lsri,\Asai,\Agliii) &=
  2 \times (h)^{\rmii{EE}} 
  = 0
  \;.
\end{align}

The magnetic contribution $(f)^{\rmii{BB}}$ is strategically analogous to
its electric counterpart. 
The $z$-integration over the single $A_z$ drawn from the Wilson line yields 
\begin{align}
\label{f:BB}
\ToptWM(\Lsri,\Agliii,\Agliii,\Lgliii) &=
  2 \times (f)^{\rmii{BB}}
  \nn &=
  - 2\gsq\CA^{2}\CF^{ }
  \int_{\vec{pk}} e^{ip_{z}L}\,
  \frac{1}{
    p^2 k^2 (p-k)^2}
  \nn &\hp{{}=-(d-1)}\times
  \Bigl[
    \mathbb{P}\,\frac{p_{z} + k_{z}}{p_{z} - k_{z}}
    \frac{1}{2}
    \bigl(
        p^2 + k^2 + (p-k)^2
      + 2(d-2)p_{z}k_{z}
    \bigr)
  - \bigl(p^2 - k^2\bigr)
  \Bigr]
  \;.
\end{align}
The corresponding master integrals of the 
Fourier transform are 
\begin{equation}    
\label{eq:master:fB}
\mathcal{I}_{\alpha\beta\gamma}^{ab} =
  \int_{\vec{pk}} e^{i p_z L}\,
 \mathbb{P}\, \frac{p_{z} + k_{z}}{p_{z} - k_{z}}
  \frac{
    p_{z}^{a}k_{z}^{b}
    }{
    [p^2 ]^{\alpha}
    [k^2 ]^{\beta}
    (p-k)^{2\gamma}}
  \;,
\end{equation}
such that
\begin{align}
\label{eq:master:f3}
  \mathcal{I}_{111}^{11} &=\frac{3}{2(4\pi L)^2}
  \;, \\
\label{eq:master:f4}
  \mathcal{I}_{011}^{00} &=
  \Bigl(\frac{\bmu^2 e^{\gammaE}}{4\pi}\Bigr)^{2\epsilon}
  \frac{ (3 d-7)
  \Gamma\left(\frac{d}{2}-1\right)
  \Gamma (d-3)}{2^{d+1}\pi ^{d-1/2}
  \Gamma\left(\frac{d-1}{2}\right)[L^2]^{d-2} }=
   \frac{1}{(2\pi L)^2}
   \Bigl(-\frac{1}{4\epsilon} +\frac34- \gammaE- \ln (\bmu L)\Bigr)
   \;, \\
\label{eq:master:f5}
  \mathcal{I}_{101}^{00} &= 0
  \;, \\
\label{eq:master:f6}
  \mathcal{I}_{110}^{00} &=
  \Bigl(\frac{\bmu^2 e^{\gammaE}}{4\pi}\Bigr)^{2\epsilon}
  \frac{ \cot (\pi d) \Gamma (2 d-4)}{2 (4\pi)^{d-2}
  \Gamma\left(\frac{d-1}{2}\right)^2 [L^2]^{d-2}}=
  \frac{1}{(2 \pi L)^2}
  \Bigl(-\frac{1}{4\epsilon} +1-\gammaE-\ln(\bmu L)\Bigr)
  \;.
\end{align}
The sum over the corresponding terms in eq.~\eqref{f:BB} is not finite, i.e.
\begin{align}
    \label{f:BB_final}
    2 \times (f)^{\rmii{BB}}&=
  -\gsq\CA^{2}\CF^{ }\Bigl[
    \mathcal{I}^{00}_{011}
  + \mathcal{I}^{00}_{101}
  + \mathcal{I}^{00}_{110}
  + 2(d-2)\mathcal{I}^{11}_{111}
  - 2\,\mathcal{I}_{0;000}
  \Bigr]
  \nn&=
  \frac{2\gsq\CA^{2}\CF^{ }}{(4\pi L)^2}\Bigl[
      \frac{1}{\epsilon}
    + 4\ln(\bmu L)
    + 4\gammaE
    - 4
  \Bigr]
  \;,
\end{align}
where the final term on the first line is the contribution 
of the simpler, final two terms on eq.~\eqref{f:BB} and
its master is the massless 
$\mathcal{I}_{0;000} = 1/(4\pi L)^2$ from eq.~\eqref{eq:I:a:0mm}. 
As we now show, this UV divergence is cancelled by an opposite
one from diagram $(h)^{\rmii{BB}}$, which reads
\begin{align}
\label{h:BB}
  \ToptWal(\Lsri,\Agliii,\Agliii) &=
  2 \times (h)^{\rmii{BB}}
  \nn &=
  2\gsq\CA^{2}\CF^{ }
  (d-1)
  \int_{\vec{pk}} e^{ip_{z}L}\,
  \frac{1}{
    p^2 k^2 (p-k)^2}
    \mathbb{P}\,
  \Bigl[
      \frac{p^2 k_{z}}{p_{z} - k_{z}}
    + \frac{k^2 p_{z}}{p_{z} - k_{z}}
  \Bigr]
  \nn &=\gsq\CA^{2}\CF^{ }
  \Bigl(\frac{\bmu^2 e^{\gammaE}}{4\pi}\Bigr)^{2\epsilon}
  \frac{L^{4-2 d} \Gamma (d-3)
   \Gamma (d)}{2^{2 d-3} \pi ^{d-1}\Gamma \left(\frac{d-1}{2}\right)^2}
  \nn &=
  - \frac{2\gsq\CA^{2}\CF^{ }}{(4\pi L)^2}\Bigl[
    \frac{1}{\epsilon}
  + 4\ln(\bmu L)
  + 4\gammaE
  - 3
  \Bigr]
  \;,
\end{align}
so that the combined result is finite again
\begin{equation}
\label{f:BB2}
  2 \times \bigl(
    (f)^{\rmii{BB}}
  + (h)^{\rmii{BB}}
  \bigr) =
  - \frac{2\gsq\CF^{ }\CA^2}{(4\pi L)^2}
  \;.
\end{equation}

At NLO two contributions to $\EB$ of mixed electric-magnetic correlations arise. 
One of them is
$(f)^{\rmii{EB}}$ and its mirrored partner
$(f)^{\rmii{BE}}$. 
Their result in momentum space is integrated over
the  $z'$-coordinate of the $\Phi$ field sourced
by the Wilson line
\begin{align}
\label{f:EB}
\ToptWM(\Lsri,\Agliii,\Asai,\Lsai)
&+
\ToptWM(\Lsri,\Asai,\Agliii,\Lsai)
= 2i \times (
    (f)^{\rmii{EB}}
  + (f)^{\rmii{BE}}) 
  \nn &=
  2i\gsq\CA^{2}\CF^{ }
  \int_{0}^{L}\! {\rm d}z'
  \nn&\hphantom{{}2i\gsq\CA^{2}\CF^{ }}\times
  \int_{\vec{pk}}
  \frac{e^{ip_{z}L}e^{-i(p_z - k_z)z'}}{
  p^2
  [k^2 + \mD^2]
  [(p-k)^2 + \mD^2]}
  \bigl(
    (p^2 + k^2 - (p-k)^2)k_{z} - 2k^2 p_{z}
  \bigr)
  \;.
\end{align}
The first three terms can be evaluated by using
a bare ($\mD\to 0$) subtraction scheme 
\begin{equation}
(f)^{\rmii{EB}} =
\bigr(
  (f)^{\rmii{EB}}
- (f)^{\rmii{EB}}\bigr|_{\mD\to 0}\bigr)
+ (f)^{\rmii{EB}}\bigr|_{\mD\to 0}
\;,
\end{equation}
where
the first bracketed term is evaluated in $d=3$ and
the second in $d=3-2\epsilon$ dimensional regularization.
The last term in eq.~\eqref{f:EB} splits into
a part treatable with the same scheme and
an additional finite part proportional to
\begin{align}
\label{eq:feb:md}
i\mD^{2}\int_{0}^{L}\! {\rm d}z'
&\int_{\vec{pk}}
  \frac{p_{z}
  e^{ip_{z}L}e^{-i(p_z - k_z)z'}
    }{
  p^2
  [k^2 + \mD^2]
  [(p-k)^2 + \mD^2]}
  =
  -\frac{ e^{-\mD L}}{(4\pi L)^2}(\mD L)
  \,\mbox{Shi}(\mD L)
  \;,
\end{align}
with the hyperbolic sine integral function $\mbox{Shi}(x)$
defined below eq.~\eqref{PT_EBBE_final}.
To evaluate this finite contribution,
we went to position space and
inspected its asymptotic behavior for small and large values of $\mD L$.
Thus, we could extract the analytic form of eq.~\eqref{eq:feb:md} which
agrees with the numerically integrated result.
After summing all terms in eq.~\eqref{f:EB},
we obtain
\begin{align}
\label{f:EB:2}
2i &\times (
(f)^{\rmii{EB}} +
(f)^{\rmii{BE}}) 
  \nn[2mm]&=
  - \frac{2\gsq\CA^{2}\CF^{ }}{(4\pi L)^2}\Bigr[
      \frac{1}{\epsilon}
    - 3 - 2 E_{1}(\mD L)
    + 2\ln\frac{(\bar\mu L)^2 e^{\gamma_E}}{\mD L}
    + 2e^{-\mD L}\mbox{Shi}(\mD L)
  \Bigr] 
  \;.
\end{align}
The second mixed $\EB$ contribution is
$(h)^{\rmii{EB}}$ which has the momentum space representation after
the $z$-integration of the single Wilson-line sourced field
\begin{align}
\label{h:EB}
  \ToptWal(\Lsri,\Agliii,\Asai)
&+
  \ToptWar(\Lsri,\Agliii,\Asai)
= 2i \times (
(h)^{\rmii{EB}} +
(h)^{\rmii{BE}}) 
  \nn &=
  -2i\gsq\CA^{2}\CF^{ } (d-1)
  \int_{0}^{L}\! {\rm d}z'
  \int_{\vec{pk}}
  \frac{p_z e^{ip_{z}L}}{p^2}
  \frac{e^{ik_{z}(z'-L)}}{[k^2 + \mD^2]}
  \nn &=
  -2\gsq\CA^{2}\CF^{ } (d-1)
  \int_{0}^{L}\! {\rm d}z'
    \bigl(\partial_{\rmii{$L$}}\mathcal{I}_{1;0}(L)\bigr)
    \mathcal{I}_{1;m}(L-z')
  \nn &=
  \frac{2\gsq\CA^{2}\CF^{ }}{(4\pi L)^2}\Bigl[
      \frac{1}{\epsilon}
    - 3
    - 2 E_{1}(\mD L)
    + 2\ln\frac{(\bmu L)^2 e^{\gamma_E}}{\mD L}
  \Bigr]\;.
\end{align}
Its scale dependence and $\epsilon$-poles are compensated
by diagram $(f)^{\rmii{EB}}$~\eqref{f:EB:2}.
We collect here the sum of all $EB$-terms, which is
\begin{eqnarray}
\label{ebtotal}
  2i \times (
      (f)^{\rmii{EB}}
    + (h)^{\rmii{EB}}
    + [BE]
    )
    &=&
  - \frac{4\gsq\CA^{2}\CF^{ }}{(4\pi L)^2} e^{-\mD L} \mbox{Shi}(\mD L)
  \nn &\stackrel{L\to 0}{=}& 
  -\frac{4\gsq\CA^{2}\CF^{ }}{(4\pi L)^2}\Bigl[
    \mD L - (\mD L)^2
  \Bigr]
  + \OO(L)
  \;,
\end{eqnarray}
where $[BE]$ collects mirrored contributions.

%
\subsection{Diagram $(g)$}
\label{sec:g}

The scalar electric contribution $(g)^{\rmii{EE}}$
gives rise to the following Fourier transform
\begin{eqnarray}
\label{g:EE}
\ToptWSal(\Lsri,\Asai,\Asai,\Agliii) &=&
  2 \times (g)^{\rmii{EE}}
  \nn &=&
  -2\gsq \CA^{2}\CF^{ }
  \int_{\vec{pk}} e^{ip_{z}L}
  \frac{1}{
  [p^2 + \mD^2]
  [k^2 + \mD^2]
  (p-k)^2}
  \nn &&\hp{2\times (g)^{\rmii{EE}}}\times
  \frac{1}{2}\Bigl[
      3p^2 + k^2 - (p-k)^2
    - 2(p_{z} + k_{z})p_{z}
    \Bigr]
  \;,
\end{eqnarray}
including the evaluation of the master integrals
\begin{eqnarray}
\label{gee:final}
  2 \times (g)^{\rmii{EE}} &=&
  - \frac{2\gsq\CA^{2}\CF^{ }}{(4\pi L)^2}
    \frac{e^{-\mD L}}{\mD L} \Bigl[
      2(1 + \mD L)(\ge+\ln(2\mD L)) - \mD L
  \nn && \hphantom{-\frac{2\gsq\CA^{2}\CF^{ }}{(4\pi L)^2}\frac{e^{-\mD L}}{\mD L} \Bigl[}
    + 2 e^{2\mD L} (1 - \mD L) E_{1}(2\mD L)
  \Bigr]
  \nn &\stackrel{L\to 0}{=}&
  \frac{2\gsq\CA^{2}\CF^{ }}{(4\pi L)^2}\Bigl[
    - 3 + \mD L
    - (\mD L)^2 \frac{-29 + 24\gammaE + 24 \ln (2\mD L)}{18}
  \Bigr]
  + \OO(L)
  \;.\hspace{1cm}
\end{eqnarray}

The magnetic contribution
$(g)^{\rmii{BB}}$ vanishes
in contrast to its $(g)^{\rmii{EE}}$-counterpart.
This is seen explicitly in
\begin{align}
\label{g:BB}
\ToptWSal(\Lsri,\Agliii,\Agliii,\Agliii) &=
  2 \times (g)^{\rmii{BB}}
  \nn &=
  -2\gsq\CA^{2}\CF^{ }
  \int_{\vec{pk}} e^{ip_{z}L}
    \frac{1}{p^2 k^2 (p-k)^2}
  \nn &\hp{=-2\gsq\CA^{2}\CF^{ }}\times
  \frac{1}{2}\Bigl[
     3p^2 + k^2 - (p-k)^2
    + 2(d-2) p_{z}(p_{z} + k_{z})
    \Bigr]
  \nn[1mm] &= 
  -2 \gsq\CA^{2}\CF^{ }
  \frac{3(d-3)(d-1)}{(d-4)}
  \int_{\vec{pk}} e^{ip_{z}L}
    \frac{1}{k^2 (p-k)^2}
  \;,
\end{align}
where
the last line holds in dimensional regularization and vanishes for $d=3$
since the integrand is finite.
It was derived using integration-by-parts identities at two-loop level
similar to eq.~\eqref{eq:ibp:1l}.
The relevant relation is
\begin{align}
\label{eq:ibp:2l}
  \int_{\vec{p}_{\perp}\vec{k}} \frac{p_{z}(p_{z}+k_{z})}{p^2 k^2 (p-k)^2} &=
    \frac{3}{2}
    \frac{(2d-5)}{(d-4)}\int_{\vec{p}_{\perp}\vec{k}} \frac{1}{k^2(p-k)^2}
  \;.
\end{align}

\section{Simulation details}
\label{sec:sim:details}
In order to carry out non-perturbative computations, we discretize the continuum action of EQCD \eqref{eq:EQCD_cont_action} on a three-dimensional numerical grid with lattice spacing $a$,
\begin{align}
\label{eq:EQCD_latt_action}
S_{\mathrm{EQCD},L} =& \;
  \beta \sum_{x, i>j} \Bigl( 1 - \frac{1}{3} \Box_{x,ij} \Bigr)
  \nn &
  + 2 \sum_{x,i} \Tr \, \Bigl(
      \Phi_\rmii{L}^2(x)
    - \Phi_\rmii{L}(x) U_i(x) \Phi_\rmii{L}(x + a \hat{i}) U^\dagger_i(x) \Bigr)
  \nn &
  + \sum_x \left[ Z_4 (x + \delta x) \left( \Tr \, \Phi_\rmii{L}^2(x) \right)^2 + Z_2 (y + \delta y) \Tr \, \Phi^2_\rmii{L} (x) \right]
  \\
  \Box_{x,ij} \equiv& \;
    U_i(x)
    U_j(x + a \hat{i})
    U^\dagger_i(x + a \hat{j})
    U^\dagger_j(x)
  \;,
\end{align}
where $U_i(x)$ is the gauge link in spatial direction $i$,
connecting lattice sites $x$ and $x + a\hat{i}$. 
We rescaled the adjoint scalar field to its lattice version $\Phi_\rmii{L}$ such that its wave-function normalization $Z_\Phi = 1$ is always enforced.
The subscript L will be dropped in the context of lattice calculations.
Analytical calculations in EQCD lattice perturbation theory yield expressions for
the counterterm of the inverse gauge coupling $\beta$,
the quartic coupling $\delta x$, and the
multiplicative mass ($Z_2$) and
quartic ($Z_4$) renormalization that
compensate for discretization errors up to $\OO(a)$~\cite{Moore:1997np}.
A semi-analytical computation of $\delta y$ completes the $\OO(a)$-improvement at the Lagrangian level~\cite{Moore:2019lua}. 
At the operator level, we use the lattice implementation of
the modified Wilson line~\eqref{eq:EQCD:mod:WL} described
in~\cite{Panero:2013pla}, with
the $\OO(a)$-improvement delivered in~\cite{DOnofrio:2014mld},
to connect the respective pairs of operators.
The `color-electric' field operator is discretized using 
gauge-invariant central derivative (see \cite{Moore:2020wvy}),
whereas the `color-magnetic' field operator is calculated using
the clover discretization~\cite{Weisz:1982zw,Weisz:1983bn}.
For the $\BB$-correlator, only the single-plaquette correlations of
the pair of clover operators are used, as elaborated in sec.~\ref{sec:latt}. 
Consequently, the only remaining sources of errors at $\OO(a)$ are of the form \eqref{latt_op_mix} and can be determined in an overall fit, as
outlined in sec.~\ref{sec:latt}.

Our numerical implementation is based on
{\tt openQCD-1.6} by Martin L\"uscher~\cite{openQCD}.
Using a combined update of one heatbath sweep succeeded by four over-relaxation sweeps through the volume, we update the lattice sites in a checkerboard ordering. We use the multi-level algorithm proposed by L\"uscher and Weisz~\cite{Luscher:2001up} to reduce the noise in relation to the signal for non-local operators in lattice gauge theories. In particular, we divide our volume in four sub-volumes along the $z$-axis and freeze the surfaces between them for 80 combined heatbath/over-relaxation sweeps, before we allow a single combined sweep through the entire volume.
For further details, see~\cite{Moore:2019lua,Moore:2019lgw,Moore:2020wvy}.
We give the raw data at finite lattice spacings $a$ in
an online repository~\cite{rawData}.

We provide the parameters of our simulations in tab.~\ref{sim_params}.
As argued in~\cite{Hietanen:2008tv},
EQCD possesses a mass gap and therefore finite volume effects are exponentially suppressed.
Thus, one can keep finite volume effects under control by maintaining a sufficiently large volume along the rules of thumb given in~\cite{Hietanen:2008tv}.
We measured all three correlators on the same lattice configurations introducing 
correlations among different $\gsql$ and
correlators that had to be accounted for in the jackknife error analysis by keeping
the binned jackknife data until the numerical integration and
calculating the error thereafter.
Equal computational resources were spent on all four scenarios $(T,\Nf)$;
the difference in statistics can be explained by
a slight increase in the acceptance rate for smaller $x$, since the quartic self-coupling of
the scalars is taken into account via a Metropolis step in the scalar heatbath-update.
\clearpage

\begin{table}[t] 
\centering {\footnotesize
\begin{tabular}{|c|c|c|c|c|c|}	
\hline
$\gsqa$ & $\xc$ & $\yc$ & $N_\rmii{x} N_\rmii{y} N_\rmii{z}$
& $L / a$ & statistics\\
\hline
$1/4$ & $0.08896$ & $0.452423$ & $24^2 \times 48$ & $4,6,8,10,12$ & $14260$ \\
$1/6$ & $0.08896$ & $0.452423$ & $36^2 \times 72$ & $6,12,18$ & $8440$ \\
$1/8$ & $0.08896$ & $0.452423$ & $48^2 \times 96$ & $4,6,8,12,16,20,24$ & $8709$ \\
$1/12$ & $0.08896$ & $0.452423$ & $72^2 \times 144$ & $6,12,18,24,30,36$ & $5720$ \\
$1/16$ & $0.08896$ & $0.452423$ & $96^2 \times 192$ & $4,8,12,16,24,32,40$ & $6060$ \\
$1/24$ & $0.08896$ & $0.452423$ & $144^2 \times 288$ & $6,12,18,24$ & $780$ \\
$1/32$ & $0.08896$ & $0.452423$ & $192^2 \times 384$ & $4,8,16,24,32$ & $440$ \\
\hline
$1/4$ & $0.0677528$ & $0.586204$ & $24^2 \times 48$ & $4,6,8,10,12$ & $17940$ \\
$1/6$ & $0.0677528$ & $0.586204$ & $36^2 \times 72$ & $6,12,18$ & $8700$ \\
$1/8$ & $0.0677528$ & $0.586204$ & $48^2 \times 96$ & $4,6,8,12,16,20,24$ & $8620$ \\
$1/12$ & $0.0677528$ & $0.586204$ & $72^2 \times 144$ & $6,12,18,24,30,36$ & $6820$ \\
$1/16$ & $0.0677528$ & $0.586204$ & $96^2 \times 192$ & $4,8,12,16,24,32,40$ & $6060$ \\
$1/24$ & $0.0677528$ & $0.586204$ & $144^2 \times 288$ & $6,12,18,24$ & $1400$ \\
$1/32$ & $0.0677528$ & $0.586204$ & $192^2 \times 384$ & $4,8,16,24,32$ & $820$ \\
\hline
$1/4$ & $0.0463597$ & $0.823449$ & $24^2 \times 48$ & $4,6,8,10,12$ & $18000$ \\
$1/6$ & $0.0463597$ & $0.823449$ & $36^2 \times 72$ & $6,12,18$ & $10680$ \\
$1/8$ & $0.0463597$ & $0.823449$ & $48^2 \times 96$ & $4,6,8,12,16,20,24$ & $8620$ \\
$1/12$ & $0.0463597$ & $0.823449$ & $72^2 \times 144$ & $6,12,18,24,30,36$ & $6120$ \\
$1/16$ & $0.0463597$ & $0.823449$ & $96^2 \times 192$ & $4,8,12,16,24,32,40$ & $5160$ \\
$1/24$ & $0.0463597$ & $0.823449$ & $144^2 \times 288$ & $6,12,18,24$ & $1420$ \\
$1/32$ & $0.0463597$ & $0.823449$ & $192^2 \times 384$ & $4,8,16,24,32$ & $640$ \\
\hline
$1/4$ & $0.0178626$ & $1.64668$ & $24^2 \times 48$ & $4,6,8,10,12$ & $19380$ \\
$1/6$ & $0.0178626$ & $1.64668$ & $36^2 \times 72$ & $6,12,18$ & $10680$ \\
$1/8$ & $0.0178626$ & $1.64668$ & $48^2 \times 96$ & $4,6,8,12,16,20,24$ & $8600$ \\
$1/12$ & $0.0178626$ & $1.64668$ & $72^2 \times 144$ & $6,12,18,24,30,36$ & $6500$ \\
$1/16$ & $0.0178626$ & $1.64668$ & $96^2 \times 192$ & $4,8,12,16,24,32,40$ & $3400$ \\
$1/24$ & $0.0178626$ & $1.64668$ & $144^2 \times 288$ & $6,12,18,24$ & $1300$ \\
$1/32$ & $0.0178626$ & $1.64668$ & $192^2 \times 384$ & $4,8,16,24,32$ & $680$ \\
\hline
\end{tabular} }
\caption{%
  Parameters for all EQCD multi-level simulations.
  We give the continuum expressions for
  the rescaled quartic scalar self-coupling $x_\rmii{cont}$ and
  the rescaled screening mass $y_\rmii{cont}$.
  A conversion into lattice units is done using
  the respective lattice-continuum relations~\cite{Moore:1997np,Moore:2019lua}.
}
\label{sim_params}
\end{table}

{\small
%

}
\end{document}